\documentclass[aps,nofootinbib]{revtex4}

\usepackage{amsfonts}
\usepackage{amsmath}
\usepackage{amssymb}
\usepackage{graphicx}
\usepackage{latexsym}

\def\ol{\overline}
\def\bea{\begin{eqnarray}}
\def\eea{\end{eqnarray}}
\def\beann{\begin{eqnarray*}}
\def\eeann{\end{eqnarray*}}
\def\nn{\nonumber}

\begin{document}

\title{Generalised Wick Transform in Dimensionally Reduced Gravity}

\author{Bruno Hartmann}
\email{hartmann@aei.mpg.de} \affiliation{MPI f.
Gravitationsphysik, Albert-Einstein-Institut, 14476 Golm, Germany}
\affiliation{Center for Gravitational Physics and Geometry, The
Pennsylvania State University, State College, PA 16802, USA}

\author{Jacek Wi\'sniewski}
\email{pmzjw@maths.nott.ac.uk}
\affiliation{School of Mathematical Sciences,
University of Nottingham, Nottingham, NG7 2RD, United Kingdom}
\affiliation{Center for Gravitational Physics and Geometry, The Pennsylvania State University, State College, PA 16802, USA}

\begin{abstract} In the context of canonical quantum
gravity we study an alternative real quantisation scheme, which is
arising by relating simpler Riemannian quantum theory to the more
complicated physical Lorentzian theory - the generalised Wick
transform. On the symmetry reduced models, homogeneous
Bianchi cosmology and $2+1$ gravity, we investigate its generalised
construction principle, demonstrate that the emerging quantum
theory is equivalent to the one obtained from standard
quantisation and how to obtain physical states in Lorentzian
gravity from Wick transforming solutions of Riemannian quantum
theory.
\end{abstract}

\maketitle

\section{Introduction}

The purpose of this paper is a contribution to the search
for quantum gravity, a theory which could describe the quantum
behaviour of the full gravitational field \cite{Rovelli}. This work
lies in the canonical line of research which can be characterised
by the methodology of attempting to construct a quantum theory in
which the Hilbert space carries a representation of the operators
corresponding to the metric, without having to fix a background
metric. Starting from connection dynamical formulation of general
relativity canonical quantisation successfully leads to loop
quantum gravity \cite{Abhay's Buch}. Along these lines, we analyse the
utilisation of the generalised Wick transform quantisation method
which was first proposed by Thiemann \cite{Thiemann's W-paper} and then
modified by Ashtekar \cite{Abhay's W-paper}. Its purpose is to
simplify the appearance of the last and the key step of the
canonical quantisation program, namely solving the scalar
constraint $\hat{S}_L$.

This strategy to achieve a feasible regularisation for $\hat{S}_L$
can be described as follows. We employ the idea to relate
mathematically simpler Riemannian quantum theory with the more
complicated physical Lorentzian quantum gravity by introducing a
generalised Wick transform operator $\hat{\mathbf{W}}$ on the
common kinematical Hilbert space of both theories. Classically
this strategy arose from a generalised Wick transform where
Lorentzian and Riemannian theories are related by a constraint
preserving automorphism on the algebra of $\mathbb{C}$ valued
functions on the common real phase space of these two
theories.\footnote{In quantum field theory the usual Wick
transform technique maps real phase space of Lorentzian theory to
a section in the complexified  phase space to obtain a formally
Riemannian appearance of the theory. However now one has to
incorporate complicated reality conditions to recover Lorentzian
theory. This can be avoided in our generalised Wick transform
technique.} This opens up a new avenue within quantum connection
dynamics since we will be able to find physical states in real
Lorentzian gravity by Wick transforming appropriate solutions of
the simpler Riemannian scalar constraint. To complete canonical
quantisation, we only have to consider regulating the Wick
transform operator $\hat{\mathbf{W}}$ and the simpler Riemannian
scalar constraint $\hat{S}_R$. In the classical theory the
transform is well-defined. In full quantum theory the strategy is
attractive but the transform has remained formal. To test if it
can be made rigorous and to gain insight into the resulting
physical states we will consider simpler symmetry reduced models
of $3+1$ gravity, where the issue of regularisation simplifies
considerably.

In section \ref{Kap. Bianchi models} we investigate finite dimensional
formulations of spatially homogeneous Bianchi type I and II
cosmologies. For these models one already knows what the correct
quantisation is. Thus, we will use them to test the viability of
the generalised Wick transform strategy and to learn about its
technical features. We will observe that the two procedures based
respectively on explicit Dirac quantisation of Lorentzian theory
and on its indirect quantisation due to the Wick transform program
result in identical quantum theories. Furthermore we learn that to
obtain physical states from quantum Wick transform, we have to
restrict the Hilbert space of Riemannian quantum theory to
functions which are subject to additional boundary conditions.

In section \ref{Kap.-(2+1)-Palatini} we turn to the main and most
interesting case of $2+1$ gravity. Similarly to our previous Wick
transform investigation on Bianchi models, one already knows the
resulting quantum theory for Lorentzian $2+1$ gravity. However
this quantum theory emerges in a different formulation employing
$\mathfrak{su}(1,1)$ variables, rather then in terms of
$\mathfrak{su}(2)$ variables as it will be the case from Wick
transform quantisation. In order to compare these different
quantum theories of $2+1$ gravity and thus to gain insight into
the Wick transform technique itself we will first go over to
investigate their corresponding classical formulations. In section
\ref{Kap.-(2+1)-Palatini} we systematically construct different
canonical formulations of $2+1$ dimensional classical general
relativity in connection variables and emphasise their
practicality for the purpose of canonical quantisation. We will
clarify their interplay in sections \ref{Kap. $(2+1)$ Wick
transform} and \ref{Kap. $(2+1)$ Wick rotation} by constructing
\textit{internal} phase space maps between them\footnote{Maps,
where the complexification is only performed on the fibres of our
vector bundle formulation meanwhile the base manifold remains
unchanged.} - which we call \emph{Wick rotation} and \emph{Wick
transform}. They naturally relate Euclidean and
Lorentzian formulations of classical $2+1$ gravity. We recover one
of them - the Wick transform - as the $2+1$ dimensional analog of
our generalised Wick transform in $3+1$ gravity. The other
internal phase space map - the Wick rotation - on the classical
level also relates $\mathfrak{su}(2)$ Riemannian theory to
Lorentzian theory, this time however to its $\mathfrak{su}(1,1)$
appearance.

This new classical perspective of having different classically
equivalent formulations of $2+1$ gravity allows considering
different implementations of the corresponding quantum Wick
transform program. Our main interest is to study the
appearance of the Wick transform operator $\hat{\mathbf{W}}$ as
defined in \cite{Thiemann's W-paper,Abhay's W-paper},
since that represents the general situation which we also
encounter for the Wick transform in $3+1$ gravity. In $2+1$
gravity its appearance follows from the particular formulation of
classical Lorentzian theory in terms of $\mathfrak{su}(2)$
connections. The direct way to test the corresponding Wick
transform quantisation strategy in its $\mathfrak{su}(2)$
formulation will cause, however, regularisation problems and shall be
avoided in our further discussion. The simpler framework of
$\mathfrak{su}(1,1)$, $2+1$ gravity instead allows to choose an
indirect way to test the Wick transform quantisation program. For
this purpose, we will construct the \mbox{\emph{Wick rotation}} -
a preliminary technique for our main object of interest the Wick
transform in $2+1$ gravity.

In section \ref{Kap. $(2+1)$ Wick rotation} we construct on the
classical level a phase space map from Riemannian to Lorentzian
theory. In contrast to the generalised Wick transform, where we
employ an automorphism on the algebra of phase space functions on
the common phase space of both theories, we will now successfully
construct a true isomorphism from the phase space of Riemannian
theory to the phase space of Lorentzian theory, referred to as
internal Wick rotation. Here we approach the main advantage of our
alternative construction: (i) one already knows the classical
solution space - the cotangent bundle over the moduli space of
flat $\mathfrak{su}(2)$, respectively $\mathfrak{su}(1,1)$,
connections and (ii) explicit Dirac quantisation of employing
operator constraints is equivalent to simpler reduced phase space
quantisation. Furthermore, our construction of the internal Wick
rotation can be induced to the reduced phase space, thus providing
a natural map from "physical" states of Riemannian quantum theory
to physical states of Lorentzian quantum theory. We realise that
those internal phase-space maps can not be used to regularise the
Wick transform operator $\hat{\mathbf{W}}$ geometrically. However,
their importance lies in the possibility to construct the Wick
rotation, a Wick transform for $2+1$ gravity, which characterizes
the resulting physical states from Wick transform quantisation
indirectly. In the case of space-time topology $M^{2,1}\cong
T^2\times\mathbb{R}$ (where $T^2$ is a two-torus) we will study
the resulting physical states explicitly.
Here we realise that when applied to quantum theory the Wick
transform leads us to the most interesting, space-like sector of
Lorentzian $2+1$ gravity. Moreover, it turns out that this
construction provides a dense subset of the Lorentzian physical
Hilbert space.

For a reader who would prefer a self-contained, pedagogical
introduction to the subject we recommend \cite{Bruno thesis}
before reading this paper.

\section{Generalized Wick transform for Gravity} \label{orig. W-Konzept von
Abhay}

The aim of this section is to briefly review the concept of generalized Wick
transform and the quantisation strategy proposed in
\cite{Thiemann's W-paper,Abhay's W-paper}. Euclidean canonical general relativity expressed in
terms of $SU(2)$ connection variables has simple polynomial
constraints. Lorentzian theory can be written in terms of the same
variables, but the constraints have a complicated form. Thus, the
hope is to construct an operator which would allow to map the
quantum states of the simpler, Euclidean theory into the Hilbert
space of the Lorentzian theory.

First, for motivational purpose, we define an internal phase space
isomorphism $\phi$ on the \emph{complexified} phase space of
$\mathfrak{su}(2)$ Riemannian $3+1$ Palatini theory
\begin{align}
   E^{I}_{a} & \mapsto i E^{I}_{a} \nn \\
   K^{I}_{a} & \mapsto - i K^{I}_{a} \nn
\end{align}
Here, $(E^{a}_{I},K^{I}_{a})$ are the ADM variables (triad and extrinsic
curvature) coordinatising the $\mathfrak{su}(2)$ phase-space.
Then, we lift $\phi$ to an automorphism on the algebra of $\mathbb{C}$
valued functions on the common \emph{real} phase space of
$\mathfrak{su}(2)$ Riemannian and Lorentzian theory. The result is
the following map $\mathbf{W}$, called also the (classical)
generalised Wick transform,
 \begin{equation} \label{klass. verallg. W-Trafo}
   f \;\;\; \mapsto \;\;\; \mathbf{W} (f) \; = \; f + \{f,T\} +
   \frac{1}{2!}\{\{f,T\},T\} + \ldots \; = \;
   \sum^{\infty}_{n=0}\frac{1}{n!}\{f,T\}_n
 \end{equation}
where $T$ is a complex-valued functional given by
 \begin{equation}
\label{T-generator function}
   T := \frac{i\pi}{2}\int_{\Sigma} d^3 x K^I_a E^a_I
\end{equation}
$\mathbf{W}$ is a Poisson bracket-preserving automorphism, it does
not commute however with the complex conjugation. The important
property of Wick transform from the point of view of the
quantisation is that, when applied to Riemannian constraints, it
produces the appropriate Lorentzian constraints
 \begin{equation}
 \begin{array}{cccc} \label{klassische Konstraint-Abb.}
   \mathbf{W}\:\mathcal{G}_{I} = \mathcal{G}_{I}\; , &
   \mathbf{W}\:\mathcal{V}_{b} = \mathcal{V}_{b} & \textrm{and} &
   \mathbf{W}\:\mathcal{S}_{R} = -\: \mathcal{S}_{L}
 \end{array}
 \end{equation}
Here, $\mathcal{G}_{I}$, $\mathcal{V}_{b}$ are the Gauss and vector
constraints, which have the same form in the two theories. It is only the
scalar constraints that are different and therefore we call them
$\mathcal{S}_{R}$ and $\mathcal{S}_{L}$ respectively.

 Let us now rewrite all constraints in terms of
canonically conjugate pair of (connection) variables
$(A^{I}_{a},E^{a}_{I})$, which are related to the previous ones
via $A^{I}_{a}=\Gamma^{I}_{a}-K^{I}_{a}$ (here $\Gamma$ is the
connection compatible with the triad). The Riemannian constraints
become at most second order polynomials in basic variables and
therefore Riemannian quantum theory is (relatively) easy to solve.
Assume we have done that, we can now define the quantum operator
corresponding to the Lorentzian scalar constraint using the
relation (\ref{klassische Konstraint-Abb.}). Then, a simple
calculation shows that:
 \bea
   -\; \hat{\mathcal{S}}_{L}\; := \;
   \widehat{\mathbf{W} \mathcal{S}_{R}}
   & = & \sum^{\infty}_{n=0}\; \frac{1}{n!}\;
   {\left(\frac{1}{i\hbar}\right)}^n\;
   [\;\hat{\mathcal{S}_{R}}\; ,\:
   \hat{T}\; ]_n \nn \\
    & = & exp\; {-\frac{\hat{T}}{i\hbar}}\; \circ \;
    \hat{\mathcal{S}_{R}}\;\circ\;exp\;{\frac{\hat{T}}{i\hbar}}\;\;=:\;\;
    \hat{\mathbf{W}} \; \circ \; \hat{\mathcal{S}_{R}} \; \circ
    \; \hat{\mathbf{W}}^{-1} \label{QM-W-Trafo-Herleitung}
\eea Here $\hat{T}$ is the operator version of our generating
function $T$. In the kinematical Hilbert space $\mathcal{H}_{kin}$
physical quantum states $|\psi \rangle_L $, which are defined by
$\hat{\mathcal{S}}_{L}|\psi \rangle_L := \hat{\mathbf{W}} \circ
\hat{\mathcal{S}_{R}} (\hat{\mathbf{W}}^{-1}|\psi \rangle_L) = 0$,
can now be obtained from Wick transforming the kernel of
$\hat{\mathcal{S}_{R}}$. Thus the availability of the Wick
transform could provide considerable technical simplification: the
problem of finding solutions to all quantum constraints is reduced
essentially to that of regulating relatively simple operators
$\hat{\mathcal{S}}_{R}$ and $\hat{T}$ - in contrast to
regularizing the classically more complicated Lorentzian scalar
constraint.

For our \emph{explicit} investigation of the Wick transform
quantisation strategy in symmetry reduced models of $3+1$ gravity
we will address the kinematical question: Do the constraints
$\hat{\mathcal{S}}^{(Wick)}_{L}$ from Wick transform quantisation
and $\hat{\mathcal{S}}^{(Dirac)}_{L}$ from standard Dirac
quantisation lead to identical quantum theories on
$\mathcal{H}_{kin}$? And then the dynamical question: What are the
physical states $\hat{\mathbf{W}}|\psi \rangle_R\in
\mathcal{H}_{phys}$, if $|\psi \rangle_R$ is a solution to the
Riemannian constraints?

\section{Bianchi models} \label{Kap. Bianchi models}

In full quantum theory the Wick transform strategy is attractive
but the $\hat{\mathbf{W}}$ operator is only defined formally. Our
goal is to understand subtleties in making these formal
considerations precise. Bianchi models provide a natural arena
since in this case one can write out the operators explicitly. In
these cases one already knows what the correct answer is. The
question is if the Wick transform strategy can be implemented in
detail and if it reproduces the correct answer. Thus, the purpose
of this section is to use Bianchi models to test the viability of
the generalised Wick transform strategy and to learn about its
technical features.

In the quantisation strategy of \cite{Abhay's W-paper} we work
exclusively in the real phase space with real constraint
functionals. Thus the generalised Wick transform can be used also
in equivalent geometrodynamical formulation of $3+1$ gravity.
Therein we will study symmetry reduced Bianchi cosmology as a
first test model.

To start systematically let us recall the geometrodynamical ADM
framework. For space-like foliations $\overset{3,1}{}M\cong\overset{3}{}\Sigma
\times \mathbb{R}$ the phase space of our Hamiltonian formulation
is coordinatised by the canonical pair $(q_{ab}, p^{ab})$ with
induced metric $q_{ab}$ to the spatial slice and its conjugate
momentum $p^{ab}:=\sqrt{q}(K^{ab}-K q^{ab})$ which is defined by
the extrinsic curvature $K^{ab}$. The Hamiltonian is just a sum of
first class constraints. The vector constraint
\begin{equation}
\mathcal{V}_a = q_{ab} D_c p^{bc}
\end{equation}
is independent of the signature. Lorentzian and Riemannian
theory differ only in their scalar constraint
\bea
\mathcal{S}_L = + \left(\frac{p^2}{2}-p_{ab} p^{ab}\right) + q R &
\textrm{and} & \mathcal{S}_R = - \left(\frac{p^2}{2}-p_{ab} p^{ab}\right) + q R
\eea
The equivalence of the Hamiltonian formulations of $3+1$
Palatini theory and standard  Einstein-Hilbert theory allows us to
translate the classical Wick transform concept from triads to ADM
variables. The induced generator for our generalised Wick
transform is now given by
\begin{equation} \label{ADM-T-generator}
   T = \frac{i \pi}{2} \int_{\overset{3}{}\Sigma} d^3 x \; q_{ab} \: p^{ab}
\end{equation}
Employing $\mathbf{W}$ as an automorphism on the algebra of
phase space functions will again map Riemannian to Lorentzian
constraints on the common real phase space of both theories.

In order to simplify the Wick transform quantisation program in
terms of ADM \mbox{variables} we will now investigate spatially
homogeneous cosmologies as truncated models of \mbox{$3+1$}
gravity. Classically they admit additional symmetries which will
simplify the task of quantisation considerably. Spatially
homogeneous space-times or \textit{Bianchi cosmologies} have a
$3$-dimensional isometry group $G$ acting simply transitively on a
$1$-parameter family of space-like hypersurfaces and hence provide
a natural foliation $M \cong G \times \mathbb{R}$ for our
Hamiltonian formulation. Therein each hypersurface is isometric to
$G$ equipped with a left invariant Riemannian metric. We will
further restrict to \textit{diagonal}, spatially homogeneous models
where the $4$-metric expressed in terms of left invariant
$1$-forms $\{\omega^{i}_a \}_{i=1,2,3}$ can be written as
\begin{equation}
\label{Bianchi-diag.-metric}
   g_{ab} = - N(t)^2 (dt)_{a} \otimes (dt)_{b} + \sum_{i=1}^3 q_i(t)
   \omega^i_a \otimes \omega^i_b
\end{equation}
with purely time dependent lapse function $N(t)$ and diagonal
components $q_i(t)>0$ of the metric.

We use the ADM framework above to arrive at the Hamiltonian
formulation for Bianchi models. For \textit{class A} Bianchi
models \footnote{Bianchi models are classified by the Lie algebras
corresponding to the isometry group. The space-time belongs to
\textit{class A} Bianchi models if the trace ${C^a}_{ba}$ of the
structure constants ${C^a}_{bc}$ of the Lie algebra vanishes.}
homogeneity and diagonal form of the induced metric $q_{ab}$ are
compatible with the vacuum field equations since the vector
constraint is identically satisfied. Hence we obtain a symmetry
reduced ADM formulation which will be the point of departure for
canonical quantisation. The reduced configuration space (referred
to as minisuperspace) is $3$-dimensional and parameterised by the
components of the diagonal spatial metric $q_i > 0$. The phase
space is the cotangent bundle over it with conjugate momenta
denoted by $p^i$. From the ADM formulation we obtain the induced
Poisson bracket relations as $\{q_i,p^j\} = \delta_i^j$. Only the
reduced scalar constraint remains to be imposed in Lorentzian and
Riemannian theory \begin{align}
   \mathcal{S}_L &= + \left(\frac{{(q_i p^i)}^2}{2}-
   \sum_{i}({q_i})^2 ({p^i})^2
   \right) + q R \\
   \mathcal{S}_R &= - \left(\frac{{(q_i p^i)}^2}{2}-
   \sum_{i}({q_i})^2 ({p^i})^2\right) + q R
\end{align} with $q=\det q_{ab}=q_1q_2q_3$ and scalar curvature $R$ of
our diagonal spatial metric $q_{ab}$. The symmetry reduced Wick
transform generator (\ref{ADM-T-generator}) simplifies to
\begin{equation}
\label{Bianchi-T-generator}
   T = \frac{i\pi}{2} q_i p^i
\end{equation}

Misner has introduced a useful parameterisation of the diagonal
spatial metric. Incorporating it into a canonical transformation
$(q,p) \mapsto (\beta,\pi ):=(\ln q,\: q p)$ will further simplify
the appearance of our phase space description. In the new
canonical variables the non-holonomic constraint $q_i>0$ is
automatically satisfied - we have trivial phase space topology.
Here the scalar constraint of geometrodynamics and our Wick
transform generator appear as
\begin{align}
   \mathcal{S}_L & = + \left(\frac{(\pi_1+\pi_2+\pi_3)^2}{2} -
   ({\pi_1}^2+{\pi_2}^2+{\pi_3}^2) \right) + q
   R \label{Bianchi-Misner-L} \\
   \mathcal{S}_R & = - \left(\frac{(\pi_1+\pi_2+\pi_3)^2}{2} -
   ({\pi_1}^2+{\pi_2}^2+{\pi_3}^2)\right) + q
   R \label{Bianchi-Misner-R} \\
   T & = \frac{i\pi}{2} \; (\pi_1+\pi_2+\pi_3) \label{Bianchi-T-Misner}
\end{align}
Let us remark here that, because of this simple form of $T$, in
quantum theory there will be no factor ordering problems and one
should be able to define a self-adjoint $\hat{\mathbf{W}}$. In the
Misner variables the scalar constraint has the familiar form of a
sum of a kinetic term and a potential term $q R$, which depends on
the specific Bianchi model under consideration.\footnote{Here an
additional canonical transformation \cite{Tate Uggla} would remove
the potential term and lead to a Hamiltonian description of a
massless, relativistic particle moving in a Minkowski space where
the momenta of the particle are subject to non-holonomic
constraints. For Riemannian and Lorentzian Bianchi models
different canonical transformations would lead to different phase
spaces. However for explicit Wick transform investigation we need
a common classical phase space.} We will use our Misner
parameterisation as the point of departure for canonical
quantisation of Bianchi I and Bianchi II cosmology in the next
sections.

\subsection{Wick transform for Bianchi I} \label{Kap. Bianch I}

We now want to exploit the simplicity of the scalar constraint
(\ref{Bianchi-Misner-L}, \ref{Bianchi-Misner-R}) in Bianchi I
cosmology. As mentioned before, Bianchi models are characterised
by their Lie algebras. For Bianchi I the structure constants
vanish completely. Thus the spatial, homogeneous, diagonal metric
is flat and the potential term in the scalar constraint disappears
entirely. We find identical classical phase space description for
Riemannian and Lorentzian theory, since $\mathcal{S}_R = -
\mathcal{S}_L$.

To evaluate our quantum Wick transform program we have to quantise
Bianchi I cosmology by employing the Dirac procedure of imposing
operator constraints to select the physical states.\footnote{The
goal is to represent the Poisson algebra of real classical
observables (\textit{Dirac observables} - functions on the
kinematical phase space whose Poisson brackets with the
constraints vanish weakly) by an algebra of self-adjoint
''measurement'' operators on a Hilbert space.} This can be
achieved by performing the following steps

(i) quantise the kinematical phase space, i.e. represent the
Poisson algebra of kinematical phase space functions by an algebra
of operators on the kinematical Hilbert space $\mathcal{H}_{kin}$.

(ii) restrict to the physical Hilbert space $\mathcal{H}_{phys}
\subset \mathcal{H}_{kin}$ which is defined as the solution space
to the quantum constraints. All quantised Dirac observables leave
that subspace invariant and can therefore be reduced to
$\mathcal{H}_{phys}$.

(iii) select a scalar product on the space of physical states by
demanding that quantised physical observables be promoted to
self adjoint operators on $\mathcal{H}_{phys}$.
This can be indirectly achieved by representing all Dirac
observables (what includes the set of physical observables) - i.e.
all operators which leave $\mathcal{H}_{phys}$ invariant - by self
adjoint operators. Here it is sufficient to impose that reality
condition on a complete set of generators on the algebra of Dirac
observables.

The simplicity of Bianchi I theory allows us to find an explicit
solution space to the quantum constraints as well as a complete
set of Dirac observables and thus to carry out the
quantisation program to completion.
We start to quantise our constraint system in the Misner phase
space parameterisation (\ref{Bianchi-Misner-L},
\ref{Bianchi-Misner-R}) and implement the program step by step.

We first quantise the kinematical phase space which is
topologically $\mathbb{R}^6$ and parameterised by the canonical
pair $(\beta_i, \pi_i)$. In the passage to quantum theory lets
consider $\mathcal{H}_{kin} = L^2(\mathbb{R}^3)$ to arise from the
space of distributions $\psi$ over the configuration space.
Thereon we represent our Poisson algebra of canonical pairs
$(\beta_i, \pi_i)$ by an algebra of configuration operators
\hbox{$\hat{\beta_i}:= \beta_i\cdot \:$} and momentum operators
$\hat{\pi_i}:=-i\hbar \partial / \partial \beta_i$.

The next task is to single out the physical Hilbert space
$\mathcal{H}_{phys}$ by solving the quantum constraint
\begin{equation}
\label{QM-Bianchi-Misner-LR}
   - \hat{\mathcal{S}}_L = \hat{\mathcal{S}}_R = \hbar^2 \left[
   \frac{1}{2}\left(\frac{\partial}{\partial\beta_1}
   +\frac{\partial}{\partial\beta_2}+\frac{\partial}{\partial\beta_3}\right)^2
   - \left(\frac{\partial^2}{{\partial\beta_1}^2}+\frac{\partial^2}
   {{\partial\beta_2}^2}+
   \frac{\partial^2}{{\partial\beta_3}^2}\right) \right]
\end{equation}
We can diagonalise $\hat{\mathcal{S}}_R$ by employing a linear
coordinate transformation on the configuration space
\begin{equation}
   \left(\begin{array}{c} \beta_0 \\ \beta_+ \\ \beta_- \end{array}\right)
   := \frac{1}{3}
   \left(\begin{array}{ccc} 1&1&1 \\ -1&-1&2 \\ -1&2&-1 \end{array}\right)
   \left(\begin{array}{c} \beta_1 \\ \beta_2 \\ \beta_3 \end{array}\right)
\end{equation}
Finally we obtain our quantised scalar constraint
(\ref{QM-Bianchi-Misner-LR}) and Wick transform generator
(\ref{Bianchi-T-Misner}) as
\begin{align}
   \hat{\mathcal{S}}_R & = - \hat{\mathcal{S}}_L = \: - \: \frac{3}{2}\:\hbar^2
   \left(- \frac{\partial^2}{{\partial\beta_0}^2}
   + \frac{\partial^2}{{\partial\beta_+}^2}+
   \frac{\partial^2}{{\partial\beta_-}^2}\right) = - \:\frac{3}{2}\:\hbar^2\;
   \Box \label{QM-Bianchi-Misner-LR-b} \\
   \hat{T} & = \: \frac{3}{2} \: \pi \hbar \; \frac{\partial}{\partial
   \beta_0}\label{T-gen-I-b}
\end{align}
Thus our space of physical states $\mathcal{H}_{phys}$ will arise
from solutions to the quantum constraint equation $ \Box \: \psi =
0 $. Or, after employing a Fourier transform, in the unitarily
equivalent momentum representation from solutions of $(-{\pi_0}^2
+ {\pi_+}^2 + {\pi_-}^2 ) \: \varphi = 0$, which are spanned by
states of the form:
\begin{equation}
  \label{rigging map states}
   \varphi_{phys}(\pi):=\delta(\pi_0-\sqrt{{\pi_+}^2 + {\pi_-}^2
   })\cdot \varphi(\widetilde{\pi})
\end{equation}
where $\varphi_{phys}$ is a distribution on the kinematical
configuration space $\mathbb{R}^3$ and $\varphi\in L^2(\mathcal{L})$
a square integrable function on the light cone
$\mathcal{L}\subset\mathbb{R}^3$. From the simple scalar
constraint, a function of $\pi_{i}$ alone, we obtain a common
pre-factor $\delta(\pi_0-\sqrt{{\pi_+}^2 + {\pi_-}^2 })$ to all
distributional solutions. Hence each state $\varphi_{phys}$ is
completely characterised by functions $\varphi\in L^2(\mathcal{L})$
on the light-cone, which shall be used from now on to represent
elements in $\mathcal{H}_{phys}$.

The next step in our quantisation procedure is to find a complete
set of Dirac observables. In momentum representation a complete
set of differential operators on $L^2(\mathbb{R}^3)$ which leaves
$L^2(\mathcal{L})$ invariant can contain arbitrary
configuration operators - here we select operator analogs to the
functions $(\pi_+,\pi_-)$. However the vector fields corresponding to
the momentum operators therein are restricted to lie in the
tangent space $T\mathcal{L}$. We find a complete set of them from
the canonical immersion
\begin{equation}
\begin{array}{cccc} \label{Licht-Kegel-Immersion}
   \phi : & \mathbb{R}^2 & \rightarrow & \mathcal{L} \;\;\;\; \subset
   \;\;\;\; \mathbb{R}^3 \\
   & (\tilde{\pi}_+, \tilde{\pi}_-) & \mapsto & (\sqrt{{\tilde{\pi}_+}^2 +
   {\tilde{\pi}_-}^2}, \tilde{\pi}_+ , \tilde{\pi}_- )
\end{array}
\end{equation}
of the light cone into the kinematical configuration space.
Here we choose momentum \mbox{operators} which are slightly
different from the canonical base vector fields \bea
   \hat{v}_+ & := & \pi_0 \cdot d\phi \left(
   \frac{\partial}{\partial\tilde{\pi}_+}
   \right) = \pi_0 \frac{\partial}{\partial \pi_+} + \pi_+ \frac{\partial}{\partial
   \pi_0} \\
   \hat{v}_- & := & \pi_0 \cdot d\phi \left(
   \frac{\partial}{\partial\tilde{\pi}_-}
   \right) = \pi_0 \frac{\partial}{\partial \pi_-} + \pi_- \frac{\partial}{\partial
   \pi_0}
\eea The set of operators $(\hat{\pi}_+,\hat{\pi}_-,\hat{v}_+,
\hat{v}_-)$ obviously generates the entire algebra of Dirac
observables on $\mathcal{H}_{phys}$ (in momentum representation).
A straightforward computation of their Lie brackets shows, that
the above set of generators enlarged by the two algebraically
related operators $\hat{v}_0 := [ \hat{v}_+,\hat{v}_- ] = - \pi_-
\frac{\partial}{\partial \pi_+} + \pi_+ \frac{\partial}{\partial \pi_-}$
and $\hat{\pi}_0 = [ \hat{v}_\pm ,\hat{\pi}_\pm ]$ forms a Lie algebra
which is isomorphic to the Lie algebra of the Poincare group in
3-dimensional Minkowski space.\footnote{Here the configuration
operators $\hat{\pi}_I$ generate translations while the momentum
operators $\hat{v}_I$ generate Lorentz transformations.}

We complete our quantisation procedure by specifying an inner
product on $\mathcal{H}_{phys} = L^2(\mathcal{L}, \mu)$ in a way
that Dirac operators corresponding to real physical observables
are self adjoint. Here it is sufficient to impose that reality
condition to our complete set of generators on the algebra of
Dirac observables. For any two physical states $\varphi ,\psi \in
\mathcal{H}_{phys}$ our inner product
\begin{equation}
   \langle \varphi , \psi \rangle = \int_{\mathcal{L}} \bar{\varphi}\:\psi\;\mu
\end{equation}
is determined by the measure $\mu$ on the light cone
$\mathcal{L}$. Regarding our choice of momentum operators
$\hat{v}_I$ we find the most convenient measure on
$L^2(\mathcal{L}, \mu)$ by employing our immersion
$(\ref{Licht-Kegel-Immersion})$ to the standard flat measure on
$\mathbb{R}^{3}$ to get $\phi^\ast \mu_0 =
\frac{1}{\pi_0} \: d\tilde{\pi}_+ \wedge d\tilde{\pi}_- $ with respect
to which the pulled back vector fields $v_I$ are divergence-free.
Finally we approach our desired goal and prove their corresponding
momentum operators $\hat{v}_I$ to be self adjoint on
$L^2(\mathcal{L}, \;\mu_0 |_{\mathcal{L}}\: := \frac{1}{\pi_0} \:
d\pi_+ \wedge d\pi_-)$.

This completes the implementation of Dirac's quantisation program
for the Riemannian and Lorentzian Bianchi type I model. In the
final picture we describe the Hilbert space of physical states in
its momentum representation by $L^2(\mathcal{L}, \;\mu_0 )$.
Thereon we find the Wick transform generator (\ref{T-gen-I-b}) in
the simplified form
\begin{equation}
   \hat{T} = \:- \:i\: \frac{3}{2} \: \pi \hbar \; \pi_0 \label{T-gen-I-c}
\end{equation}
Thus $\hat{T}$ is anti self-adjoint, whence $\mathbf{\hat{W}}$
is self-adjoint.

Now we have everything together to test our quantum Wick transform
concept in the symmetry reduced Bianchi I model. We will
explicitly compare the two quantum theories based respectively on
Wick transform quantisation and on the previously introduced Dirac
procedure. In the degenerate case of Bianchi type I  we will find
both schemes to be equivalent.

The idea was to solve the scalar constraint for Lorentzian gravity
by Wick transforming solutions of the scalar constraint of
Riemannian theory. The first issue to discuss was the
regularisation of the corresponding Wick transform operator
$\hat{\mathbf{W}}$. Due to the simple appearance of its generator
$\hat{T}$ the Wick transform operator is obviously well defined
\begin{equation} \label{Bianchi-QM-W-Trafo}
   \hat{\mathbf{W}} = \exp\left( - \frac{1}{i\hbar} \hat{T} \right)
   = \exp\left( \frac{3}{2}\: \pi \:\hat{\pi}_0 \right)
\end{equation}

On the kinematical Hilbert space (in momentum representation) we
further realise that both Lorentzian scalar constraint operators
which arise from Wick transform quantisation
or from Dirac's procedure in
(\ref{QM-Bianchi-Misner-LR-b}) are identical
\begin{align}
   & \hat{\mathcal{S}}_{L}^{(\textrm{Wick})} :=  - \hat{\mathbf{W}} \;
   \circ \; \hat{\mathcal{S}_{R}} \; \circ
      \; \hat{\mathbf{W}}^{-1} \\
   &  =  - \exp\left( \frac{3}{2} \pi \hat{\pi}_0
      \right)\cdot \frac{(-3)}{2} \hbar^2 (-{\pi_0}^2 + {\pi_+}^2 +
      {\pi_-}^2) \exp\left( - \frac{3}{2} \pi \hat{\pi}_0
      \right) = \hat{\mathcal{S}}_{L}^{(\textrm{Dirac})}
\end{align}
where the Riemannian scalar constraint
(\ref{QM-Bianchi-Misner-LR-b}) is also chosen in momentum
representation.

The final step in our Wick transform quantisation program is to
introduce an inner product on the space of physical states
$\mathcal{H}_{phys}$. Following our strategy to obtain
(normalised) physical states $|\varphi_{L}\rangle :=
\hat{\mathbf{W}} |\varphi_{R}\rangle$ in Lorentzian theory from
Wick transforming solutions of the Riemannian scalar constraint,
we seek for a criterion how to select appropriate
$|\varphi_{R}\rangle$ in the solution space
$\textrm{Ker}(\hat{\mathcal{S}}_R)$ of the Riemannian scalar
constraint (with $\:\textrm{Ker}(\hat{\mathcal{S}}_R)$ described
by the space of arbitrary functions on the light cone
$\mathcal{L}$). For Bianchi I we already know $\mathcal{H}_{phys}$
from our previous Dirac quantisation method. Thus we easily find
the desired set of appropriate Riemannian solutions
$|\varphi_{R}\rangle\in\textrm{Ker}(\hat{\mathcal{S}}_R)$ from
utilising the inverse Wick transform operator as
$\hat{\mathbf{W}}^{-1} \circ \mathcal{H}_{phys} \subset
\textrm{Ker}(\hat{\mathcal{S}}_R)$. Finally we can characterise
the subspace $\hat{\mathbf{W}}^{-1} \circ \mathcal{H}_{phys}$ of
appropriate solutions to the Riemannian scalar constraint as the
space of all functions \bea
   \varphi_{R} := \varphi\cdot \exp\left( - \frac{3}{2} \pi \pi_0
   \right) & \textrm{with} & \varphi \in L^2(\mathcal{L}, \mu_0|_{\mathcal{L}}    = \frac{1}{\pi_0} \: d\pi_+ \wedge d\pi_-)
\eea We learn that to obtain (normalised) physical states from
quantum Wick transform we have to restrict to the above subspace
in the Hilbert space $L^2(\mathcal{L}, \mu_0|_{\mathcal{L}})$ of Riemannian
quantum theory. In contrast general solutions in Riemannian
quantum theory might get Wick transformed to functions which
diverge at the boundary of the configuration space.

\subsection{Wick transform for Bianchi II}

In order to canonically quantise class A Bianchi models we chose
to start classically from their parameterisation in terms of
Misner variables. Only the Riemannian or Lorentzian scalar
constraint (\ref{Bianchi-Misner-R}, \ref{Bianchi-Misner-L})
remains to be solved, where the potential term $qR$ depends on the
type of Bianchi model. The explicit common quantisation of both
Riemannian and Lorentzian Bianchi type I models was easy to
accomplish because they are flat. Now we want to study the simplest
representative for the remaining Bianchi models with non-vanishing
potential term: that is Bianchi type II.

Let us recall that Bianchi cosmologies are classified by the Lie
algebra which corresponds to their isometry group.
In class A Bianchi models structure constants are trace free and
can therefore be expressed in terms of a symmetric matrix
$n^{ab}$: ${C^a}_{bc} = \epsilon_{mbc}n^{ma}$, where
$\epsilon_{mbc}$ is the total antisymmetric symbol. In the type II
model $n^{ab}$ has signature $(0,0,+)$. Without loss of generality
we can assume diagonal form $n^{ab} =
\textrm{diag}(0,0,n)^{ab}$.\footnote{To approach a Hamiltonian
formulation in the beginning of section \ref{Kap. Bianchi models}
we choose a basis of left invariant 1-forms $\omega_a^i$ such that
the induced metric $q_{ab}$ is diagonal. By employing a
1-parameter family of orthogonal transformations we can further
achieve that, when expressed in the new basis
$\tilde{\omega}_a^i$, our symmetric matrices $q_{ab}$ and $n^{ab}$
are simultaneously diagonalised on each slice of our symmetry
induced space-time foliation.} Employing the simple structure of
the Lie algebra and starting from the diagonal form
(\ref{Bianchi-diag.-metric}) of our spatial homogeneous metric
$q_{ab}$ the scalar curvature $R$ is obtained from a
straightforward computation as
\begin{equation}
   R = - n^2 \frac{q_3}{q_1q_2}
\end{equation}
Hence we find for Bianchi type II the scalar constraints
(\ref{Bianchi-Misner-L}, \ref{Bianchi-Misner-R}) equipped with the
simple potential term $qR = -\: n^2\cdot{q_3}^2$. Expressed in the
canonically conjugate Misner variables $(\beta_i, \pi_i)$ we
obtain Lorentzian and Riemannian scalar constraint as well as the Wick
transform generator $T$ as
\begin{align}
   \mathcal{S}_L & = + \left(\frac{(\pi_1+\pi_2+\pi_3)^2}{2} -
   ({\pi_1}^2+{\pi_2}^2+{\pi_3}^2) \right)
   - \: n^2 e^{2\beta_3} \label{Bianchi-II-Misner-L} \\
   \mathcal{S}_R & = - \left(\frac{(\pi_1+\pi_2+\pi_3)^2}{2} -
   ({\pi_1}^2+{\pi_2}^2+{\pi_3}^2)\right)
   - \: n^2 e^{2\beta_3} \label{Bianchi-II-Misner-R} \\
   T & = \frac{i\pi}{2} \; (\pi_1+\pi_2+\pi_3) \label{Bianchi-II-T-Misner}
\end{align}
As in our quantum Wick transform investigation in the degenerate
Bianchi type I model in section \ref{Kap. Bianch I}, we will use
this Hamiltonian phase space description as the starting point for
quantisation \'a la Dirac.

First we quantise the kinematical phase space and obtain the same
kinematical Hilbert space $\mathcal{H}_{kin}=L^2(\mathbb{R}^3)$ as
before constrained by operators analogous to $\hat{\mathcal{S}}_R$
and $\hat{\mathcal{S}}_L$ in (\ref{QM-Bianchi-Misner-LR}) which
are supplemented with the potential term $- n^2 e^{2\beta_3}$. To
simplify the appearance of these operators we employ the same
coordinate transformation on the configuration space and obtain
the quantised scalar constraints and Wick transform generator as
\begin{align}
   \hat{\mathcal{S}}_L & = \: + \: \frac{3}{2}\: \hbar^2\; \Box \; -
   \; n^2\: e^{2\beta_0} \: e^{2\beta_+} \label{Bianchi-II-QM-L}\\
   \hat{\mathcal{S}}_R & = \: - \: \frac{3}{2}\: \hbar^2\; \Box \; -
   \; n^2\: e^{2\beta_0} \: e^{2\beta_+} \\
   \hat{T} & = \: \frac{3}{2}\: \pi\hbar \: \frac{\partial}{\partial \beta_0}
\end{align}
In contrast to Bianchi type I quantisation the solution space to
the more complicated Riemannian and Lorentzian constraint operator
is not obvious at all. Since we have no explicit expression for
the physical Hilbert space $\mathcal{H}_{phys}$ we will not
complete Dirac's quantisation procedure and therefore not Wick
transform physical states. We finish our investigation of Bianchi
type II by studying the quantum Wick transform kinematically.
Similar to its implementation on Bianchi type I, we find on
$\mathcal{H}_{kin}$ that the two formulations based respectively
on Wick transform quantisation and explicit Dirac quantisation are
equivalent.

First we study the regularisation of the corresponding Wick
transform operator. Like in our previous Bianchi I quantisation
(\ref{Bianchi-QM-W-Trafo}) we obtain
\begin{equation}
\label{Bianchi-II-QM-W-Trafo}
   \hat{\mathbf{W}} = \exp\left( - \frac{1}{i\hbar} \hat{T} \right)
   = \exp\left( i\: \frac{3}{2}\: \pi \: \frac{\partial}{\partial \beta_0} \right)
\end{equation}
as the exponential of an anti self-adjoint operator, i.e.
$\hat{\mathbf{W}}$ is well defined. On the kinematical Hilbert space
we further realise that both Lorentzian constraints $\hat{\mathcal{S}}_L^{\textrm{(Dirac)}}$
and $\hat{\mathcal{S}}_L^{\textrm{(Wick)}}$ which arise from
Dirac's procedure in (\ref{Bianchi-II-QM-L}) and, respectively, from Wick
transform quantisation are identical
\begin{align}
   \hat{\mathcal{S}}_{L}^{(\textrm{Wick})} := & - \hat{\mathbf{W}} \;
   \circ \; \hat{\mathcal{S}_{R}} \; \circ
      \; \hat{\mathbf{W}}^{-1} \\
      = & - \exp\left( i \frac{3}{2} \pi  \frac{\partial}{\partial \beta_0}
      \right)\circ \left( - \frac{3}{2} \hbar^2 \Box -
       n^2 e^{2\beta_0} e^{2\beta_+}\right) \circ \exp\left( - i \frac{3}{2}
      \pi  \frac{\partial}{\partial \beta_0} \right) \\
      = & -\left( - \frac{3}{2} \hbar^2\: \Box  \; + \;
       n^2 e^{2\beta_0} e^{2\beta_+}\right) =
       \hat{\mathcal{S}}_{L}^{(\textrm{Dirac})}
\end{align}

To summarise our investigation of the quantum Wick transform
program on Bianchi type I and II we can conclude that restricted
to these test models there is no regularisation problem for
$\hat{\mathbf{W}}$ arising, what one might expect due to its
complicated definition. Employing
canonical transformations to these symmetry reduced ADM
formulations simplified the generalised Wick transform operator
considerably. We observe that the classically equivalent
appearance of the Lorentzian scalar constraint arising from
generalised Wick transform results after canonical quantisation in
the same constraint operator
$\hat{\mathcal{S}}_{L}^{(\textrm{Wick})}$ as the one obtained from
direct canonical quantisation of the Lorentzian scalar constraint
$\hat{\mathcal{S}}_{L}^{(\textrm{Dirac})}$. Our test on Bianchi
models shows, that the generalised Wick transform can be
implemented in detail and reproduced the same quantum theory as
explicit Dirac quantisation. Furthermore we explicitly observed in
our Bianchi type I quantisation, that to obtain physical states
from quantum Wick transform, we have to restrict the Hilbert space
of Riemannian quantum theory to functions which are subject to
additional boundary conditions.

\section {$2+1$ Palatini theory} \label{Kap.-(2+1)-Palatini}

In this section we turn to the main and most interesting case of
$2+1$ gravity. Similarly to our investigation on Bianchi
models our goal is to use the symmetry reduced framework of $2+1$
gravity to study the Wick transform quantisation programme by
comparing the emerging quantum theory with known results from
standard quantisation.

In this section, we will be mainly concerned with classical $2+1$
Palatini theory and discuss canonical quantisation
later on. Here we describe different formulations of $2+1$
dimensional classical general relativity in vacuum. In our
Hamiltonian description the pair of canonically conjugate
variables is given by a connection one-form $A^{I}_{a}$ and a
densitized dyad $E^{I}_{a}$. Our descriptions will differ depending on
their gauge group choice in the vector bundle formulation.
According to the variables chosen  our theories will represent
either Riemannian or Lorentzian gravity. Furthermore we describe
two Lorentzian theories and obtain different sets of associated
constraints which are very similar to constraints encountered in
$3+1$ gravity between the $SU(2)$ and $SL(2,\mathbb{C})$
Lorentzian theories. Later, by relating these classical Lorentzian
theories to Riemannian theory, we will construct and study different
implementations of the corresponding quantum Wick transform program.

\subsection{$\mathfrak{su}(2)$ Riemannian \; and \; $\mathfrak{su}(1,1)$
Lorentzian $2+1$ theory}

\label{standard (2+1) darst.}

Let us briefly introduce the Hamiltonian formulation of $2+1$
Palatini theory following the detailed implementation and notation
in the review article by Romano \cite{Romano}. The classical Lagrangian
starting point is standard Einstein-Hilbert action in metric
variables. We introduce triads as soldering forms which provide an
isomorphism $e^I_a(p): T_pM \rightarrow \mathfrak{g}$. For
Riemannian theory we choose $\mathfrak{g}$ to be
$\mathfrak{su}(2)$ and for Lorentzian \mbox{gravity $\mathfrak{su}(1,1)$.}
Then, we find an equivalent configuration space description by
reformulating the Einstein-Hilbert action as a functional of
triads $e^I_a$ and connections $A^I_a$. We obtain the $2+1$
Palatini action as
\begin{equation} \label{(2+1)Palatini-action}
   S_P (A,e):= \frac{1}{4}\int_{M}\tilde{\eta}^{abc} e_{aI}
   F^{I}_{bc}
\end{equation} where $\tilde{\eta}^{abc}$ represents the densitized volume form
and $F^I_{ab}=2\partial_{[a}A^I_{b]}+[A_a,A_b]^I$ is the curvature
of the covariant derivative $D_a=\partial_a+[A_a,\cdot ]$ which
is determined by the connection $A^I_a$. When we vary $S_P(A,e)$
with respect to connections and triads we obtain two
Euler-Lagrange equations of motion which are again equivalent to
standard Einstein equation in metric variables.

To provide the point of departure for canonical quantisation we
will now move over to the Hamiltonian formulation. Let our base
manifold $M \cong \Sigma \times \mathbb{R}$ be foliated by the
images of a one-parameter family of embeddings $\mathcal{E}_t
:\Sigma \rightarrow \Sigma_t \subset M$. For simplicity we further
assume $\Sigma$ to be compact. Here $\mathbb{R}$ coordinates
represent evolution parameter from one $2$-dimensional $t=const.$
surface $\Sigma_t$ to the other, which does not necessarily have
the interpretation of time. This is where a conceptual difference
to $3+1$ Palatini theory arises, where we allowed space-like
foliations only. Now we use arbitrary foliations! After performing
the Legendre transform we obtain the resulting Hamilton
description as follows. The phase space consists of pairs
$(A^I_a,E^J_b)$, where $A^I_a$ is the pull back of the connection
to the $2$-dimensional slice $\Sigma$ and $E^J_b$, its conjugate
momentum, is the pull back of our triads to $\Sigma$ with density
weight one. For convenience we will call
$E^I_a$ dyad from now on. Following Dirac constraint analysis we
find, similarly as in $3+1$ Palatini theory, our Hamiltonian as a
linear combination of \textit{Gauss} and \textit{curvature
constraint} functions \bea
   \mathcal{G}_I & := & D_a E^a_I = 0 \label{Palatini-Gauss}\\
   \mathcal{F}_{ab}^{I} & := & F^I_{ab} = 0 \label{Palatini-Kruemmung}
\eea which are independent of the signature and remarkably only
linear or independent in momenta $E^I_a$. In contrast to $3+1$
gravity, where we encountered additional second class constraints, in
$2+1$ Palatini theory the constraint system is first class. Given
a test field $v^I$ the Gauss constraint Hamiltonian vector field
generates infinitesimal vector bundle gauge transformations \bea
   A^I_a & \mapsto & A^I_a - \epsilon D_a v^I \\
   E^I_a & \mapsto & E^I_a - \epsilon [v,E_a]^I
\eea which similarly arise in Yang-Mills theory. The $1$-parameter
family of diffeomorphisms generated by the curvature constraint
Hamiltonian vector field is \bea
   A^I_a & \mapsto & A^I_a \\
   E^I_a & \mapsto & E^I_a + \epsilon D_a v^I
\eea

To summarise, our constrained Hamiltonian system is parameterised by
$\mathfrak{g}$-valued connections $A^I_a$ and dyads $E^I_a$ which
obey first class constraints (\ref{Palatini-Gauss}) and
(\ref{Palatini-Kruemmung}). In our Palatini formulation Riemannian
and Lorentzian $2+1$ gravity are only distinguished by their
internal vector space $\mathfrak{su}(2)$ and
$\mathfrak{su}(1,1)$, respectively.

For non-degenerate dyads $\mathcal{F}$ can be decomposed into its
"normal" and "tangential" internal vector space component to
obtain Gauss, vector and scalar constraint which are similar to
constraints encountered in $3+1$ gravity
\begin{align}
  \mathcal{G}_{I} &= D_{a}E^{a}_{I} = 0 \\
  \mathcal{V}_{b} &= E^{a}_{I} F^{I}_{ab} = 0\\
  \mathcal{S} &= \varepsilon_{IJK} E^{Ia} E^{Jb} F^{K}_{ab} = 0
  \label{su(1,1)Lorentz-Skalarkonstraint}
\end{align}

\subsection{Lorentzian $2+1$ theory using $\mathfrak{su}(2)$}
\label{Killingreduktionsabschnitt}

So far our Lorentzian theory uses $\mathfrak{su}(1,1)$ valued
dyads and connections as phase space variables. We will now
provide an alternative formulation of Lorentzian $2+1$ gravity
using $\mathfrak{su}(2)$ valued dyads and connections, i.e.
embedded into the same phase space as Riemannian theory. To
achieve that formulation we will first understand how our present
Lorentzian $\mathfrak{su}(1,1)$ theory naturally arises by
symmetry reduction of full Lorentzian $3+1$ gravity. Using an
alternative reduction process we will finally obtain Lorentzian
$2+1$ Palatini theory in terms of $\mathfrak{su}(2)$
variables.

Let us now rediscover Lorentzian $2+1$ Palatini theory as a
suitable Killing reduction of Lorentzian $3+1$ Palatini theory.
In that case we will
assume the existence of a fixed spatial symmetry represented by a
constant Killing vector field $\xi^a$, such that our base manifold
foliates $\overset{3,1}{}M \cong \overset{2,1}{}\Sigma \times
\mathbb{R}$ with $\mathbb{R}$ representing space-like integral
curves of $\xi^a$. We decompose the Lorentzian $3+1$ Palatini
action with respect to the given symmetry
induced foliation. The result is analogous to the well known
space-time decomposition of the $3+1$ Palatini action. In order
to reduce this action to the previously discussed Lorentzian $2+1$
Palatini action (\ref{(2+1)Palatini-action}) with gauge group
$SO(2,1)$ we have to impose the following obvious conditions: (i)
gauge fix our tetrads to $( \overset{4}{}e \cdot \xi)^i = const.$
and (ii) gauge fix our connections by $\mathcal{L}_{\vec{\xi}} \;
\overset{4}{}A_a^{ij}=0$ and $(\xi \cdot
\overset{4}{}A)^{ij}=0$. Effectively we partially gauge fix our
tetrads and connections onto the section $\overset{2,1}{}\Sigma$
which arose from the foliation pursuant to the fixed spatial
symmetry.

Partially gauge fixed action can be immediately identified
with Lorentzian $2+1$ Palatini action (\ref{(2+1)Palatini-action})
where we use $\mathfrak{su}(1,1)$ phase space variables. According
to our previous discussion we move over to its Hamiltonian
description using arbitrary foliations on $\overset{2,1}{}\Sigma$
consequently without further gauge fixing. Eventually we find a
Killing reduced version of Lorentzian $3+1$ Palatini theory which
is identical to $\mathfrak{su}(1,1)$ Lorentzian $2+1$ Palatini
theory.

In the following symmetry based reduction process of Lorentzian
$3+1$ Palatini theory we again assume the existence of a fixed
spatial symmetry represented by a constant space-like Killing
vector field $\xi^a$. In particular this indirectly gauge fixes
our tetrads to be adapted to the foliation induced by $\xi^{a}$
and constant along $\xi^a$. But in
contrast to the previous reduction, where we first decomposed our
gauge restricted Palatini action with respect to the symmetry and
afterwards with respect to an arbitrary space-time foliation, we
will now start
with a suitable space-time decomposition. Consequently we first
enter the Hamiltonian formulation of $3+1$ Palatini theory where
we restrict ourselves to
space-like foliations $\overset{3,1}{}M \cong \overset{3}{}\Sigma
\times \mathbb{R}$ whereas our symmetry further restricts our
triads to be constant along $\xi^a$ in the spatial slice
$\overset{3}{}\Sigma$. That additional limitation has no influence
to the Dirac constraint analysis. Therefore the reduced phase
space of Lorentzian $3+1$ Palatini theory is parameterised
by the $\mathfrak{so}(3)$ valued pair $(E^a_I,K^I_a)$ on
$\overset{3}{}\Sigma$ and subject to Gauss, vector and scalar
constraint
\begin{align}
   \mathcal{G}_{I} &:= \varepsilon_{IJK} K^J_a E^{aK}= 0 \nn\\
   \mathcal{V}_{b} &:= 2 E^{a}_{I} D_{[a}K^I_{b]} = 0 \nn \\
   \mathcal{S}_L &:= -q R + 2 E^{[a}_I E^{b]}_J K^J_a K^I_b = 0
   \nn
\end{align}
Again the remnant of our additional spatial symmetry in
$\overset{3}{}\Sigma$ restricts our non-degenerate triads $E^a_I$
to be constant along $\xi^a$. Thus our reduced phase space
variables $(E^a_I,K^I_a)$ can be properly induced to another
foliation $\overset{3}{}\Sigma \cong \overset{2}{}\Sigma \times
\mathbb{R}$ with respect to the symmetry. They are represented by
their pull backs to $\overset{2}{}\Sigma$ and subject to the
induced constraints. We can pass back to connection dynamics by
employing canonical transformation
$(E^{a}_{I},K_{a}^{I})\mapsto(A^{I}_{a}, E^{a}_{I})$ and obtain
the associated connection dynamical description in terms of
$\mathfrak{su}(2)$-valued pairs $(A^I_a, E^a_I)$ with induced
Gauss, vector and scalar constraints
\begin{align}
   \mathcal{G}_{I} &:= D_{a}E^{a}_{I} = 0 \label{Gauss constr Lor su(2)} \\
   \mathcal{V}_{b} &:= E^{a}_{I} F^{I}_{ab} = 0 \label{Vector constr Lor su(2)} \\
   \mathcal{S}_L &:= \varepsilon_{IJK} E^{Ia} E^{Jb} F^{K}_{ab} +
   4 E^{[a}_I E^{b]}_J K^J_a K^I_b =
   0 \label{Skalar constr Lor su(2)}
\end{align}
The slightly more complicated structure of $\mathcal{S}_L$ is the
same which we encounter in $3+1$ Palatini theory during explicit
canonical transformation of the scalar constraint in ADM-variables
to the scalar constraint in connection
variables.

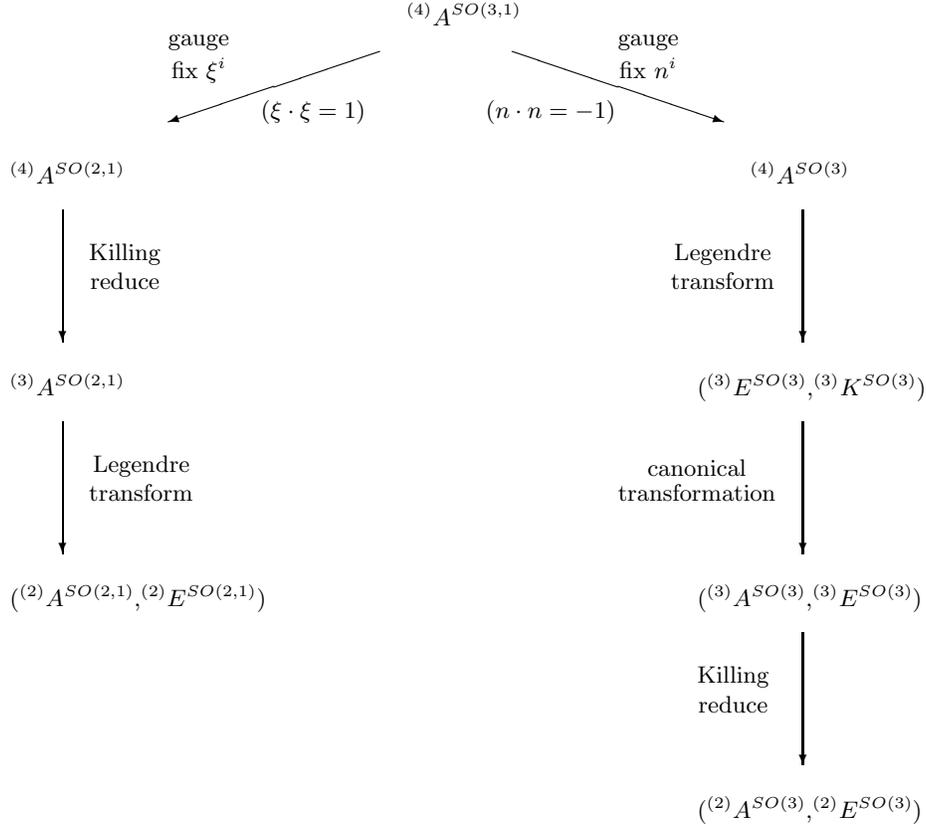
\begin{figure}
 \begin{picture}(350,350)
 \put(150,300){$^{(4)}A^{SO(3,1)}$}
 \put(140,290){\vector(-3,-1){80}}
 \put(190,290){\vector(3,-1){80}}
 \put(0,240){$^{(4)}A^{SO(2,1)}$}
 \put(280,240){$^{(4)}A^{SO(3)}$}
 \put(60,280){\shortstack{gauge \\ fix $\xi^{i}$}}
 \put(95,265){($\xi \cdot \xi = 1$)}
 \put(230,280){\shortstack{gauge \\ fix $n^{i}$}}
 \put(180,265){($n \cdot n = -1$)}
 \put(20,230){\vector(0,-1){50}}
 \put(300,230){\vector(0,-1){50}}
 \put(30,200){\shortstack{Killing \\ reduce}}
 \put(250,200){\shortstack{Legendre \\ transform}}
 \put(0,160){$^{(3)}A^{SO(2,1)}$}
 \put(260,160){($^{(3)}E^{SO(3)}$,$^{(3)}K^{SO(3)}$)}
 \put(20,150){\vector(0,-1){50}}
 \put(300,150){\vector(0,-1){50}}
 \put(30,120){\shortstack{Legendre \\ transform}}
 \put(230,120){\shortstack{canonical \\ transformation}}
 \put(0,80){($^{(2)}A^{SO(2,1)}$,$^{(2)}E^{SO(2,1)}$)}
 \put(260,80){($^{(3)}A^{SO(3)}$,$^{(3)}E^{SO(3)}$)}
 \put(300,70){\vector(0,-1){50}}
 \put(260,40){\shortstack{Killing \\ reduce}}
 \put(260,0){($^{(2)}A^{SO(3)}$,$^{(2)}E^{SO(3)}$)}
 \end{picture}
 \caption{Relations between different variables}
 \label{rys2}
 \end{figure}

Thus, performing an alternative symmetry reduction for Lorentzian
$3+1$ Palatini theory we have obtained an equivalent formulation
for Lorentzian $2+1$ Palatini theory in terms of
$\mathfrak{su}(2)$ valued connections and dyads $(A,E)$. The
constraints for those variables are more complicated than in the
$\mathfrak{su}(1,1)$ formulation. This is the price we pay for
having a compact gauge group rather than a non-compact one.
Effectively we describe Lorentzian $2+1$ theory on the same phase
space as Riemannian theory\footnote{Both appearances of Lorentzian
$2+1$ Palatini theory represent equivalent dynamics of the same
symmetric $3+1$ theory if and only if they arise from identical
dynamical and symmetry foliations $\overset{4,1}{}M\cong
\mathbb{R}_{t}\times\mathbb{R}_{\xi}\times \overset{2}{}\Sigma$.
In the second Killing reduction process we restrict to time-like
dynamical and space-like symmetry foliation. However on the first
Killing reduction process, which leads to $\mathfrak{su}(1,1)$
Lorentzian theory, we map into a phase space which encompasses
space-like dynamical foliations as well. Hence, a phase space
isomorphism, in order to relate equivalent (dynamical)
$\mathfrak{su}(2)$ and $\mathfrak{su}(1,1)$ appearances of the
same Lorentzian $2+1$ Palatini theory, should only lead to the
time-like sector in the $\mathfrak{su}(1,1)$ formulation.} subject
to the same constraints with the only exception that the scalar
constraint (\ref{su(1,1)Lorentz-Skalarkonstraint}) is replaced by
(\ref{Skalar constr Lor su(2)}). In order to further clarify the
relations between various variables discussed above we provide a
schematic description in the figure
\ref{rys2}.\footnote{Qualitative comparison of the
$\mathfrak{su}(1,1)$ and $\mathfrak{su}(2)$ constraints of the
Lorentzian theory expressed in ADM variables suggests that they
might be related by an internal phase-space isomorphism. One
could, in fact use a simple vector space isomorphism between the
two Lie algebras to induce the $\mathfrak{su}(1,1)$ Lorentzian
theory into an equivalent $\mathfrak{su}(2)$ fomrulation. This,
however, would be different from the formulation described above,
since the internal map is not (and can not be) a Lie algebra
isomorphism.} Notice also that the formulation of the Lorentzian
theory with a compact gauge group allows us to perform rigorous
constructions of various quantum operators within canonical
quantisation programme. Such constructions for the area and the
Hamiltonian operators are given in \cite{thesis}. The
constructions are analogous to the ones already known from the
$3+1$ theory.

\subsection{Natural maps between these theories} \label{viele (2+1) Abbildungen}

\begin{figure}
 \begin{center}
 \begin{picture}(310,165)
  \put(0,0){\fbox{\shortstack{Lorentzian theory \\ real $\mathbf{\mathfrak{su}(1,1)}$ \\
  $\mathcal{S}_L = \epsilon E E F =0 $}}}
  \put(0,130){\fbox{\shortstack{Lorentzian theory \\ \bf{complex} $\mathfrak{sl}(2,\mathbb{C})$ \\
  $\mathcal{S}_L = \epsilon E E F =0 $}}}
  \put(195,130){\fbox{\shortstack{Lorentzian theory \\ real $\mathfrak{su}(2)$ \\
  $\mathcal{S}_L = \epsilon E E F + \mathbf{E E K K} = 0 $}}}
  \put(217,0){\fbox{\shortstack{\bf{Riemannian} theory \\ real $\mathfrak{su}(2)$ \\
  $\mathcal{S}_R = \epsilon E E F =0 $}}}
  \put(180,15){\vector(-4,0){60}}
  \put(135,18){Wick}
  \put(130,6){rotation}
  \put(120,146){\vector(4,0){60}}
  \put(128,149){Barbero}
  \put(125,136){transform}
  \put(245,50){\vector(0,4){60}}
  \put(250,70){\shortstack{Wick \\ transform}}
 \end{picture}
 \caption{Overview of the situation}
 \label{rys}
 \end{center}
\end{figure}
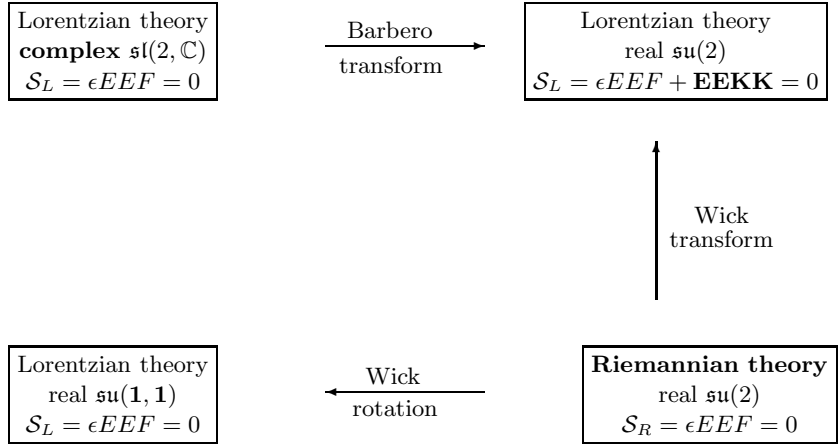

So far we have various connection dynamical formulations of $2+1$
dimensional classical general relativity. They differ (i) by their
gauge group choice in the vector bundle formulation, (ii)
represent either Riemannian or Lorentzian gravity, (iii) their
scalar constraint $\mathcal{S}$ is either simply polynomial or
difficult and (iv) they can use real or complex
connections.\footnote{In Riemannian $3+1$ Palatini theory
the canonical transformation
$(E^a_I, K^I_a) \mapsto (A^I_a :=\Gamma^I_a + K^I_a, E^a_I)$
allowed us to return to connection dynamics and simplified all
constraints. In the Lorentzian regime a similar canonical
transformation $(E^a_I, K^I_a) \mapsto (A^I_a :=\Gamma^I_a - i
K^I_a, E^a_I)$ (analog of the Barbero transform \cite{Barbero}
with parameter $\beta =-i $) leads us to self-dual
$\mathfrak{sl}(2,\mathbb{C})$ connections, providing simpler
constraint however subject to induced reality conditions. When
applied to $\mathfrak{su}(2)$ Lorentzian $2+1$ Palatini theory
this transform leads us to a similar complex
$\mathfrak{sl}(2,\mathbb{C})$ version of Lorentzian $2+1$
gravity.} From the viewpoint of canonical quantum gravity, the quantisation strategy is easier if we have compact gauge
group, polynomial constraints and work entirely with real
connections. To represent a physical theory we have to quantise
Lorentzian gravity. We realise that in all our formulations for
$2+1$ dimensional gravity introduced in the last two subsections
we have abandoned one of these simple structures. The situation is
illustrated in figure \ref{rys}.

\suppressfloats

The situation is analogous to full $3+1$ gravity. Using an
observation due to Thiemann \cite{Thiemann's W-paper}, a generalised
Wick transform allows us to circumvent that drawback and opens up
a new avenue within quantum connection dynamics. The Wick transform
strategy was to utilise a sufficiently simple, constraint preserving,
phase space map from the mathematically simpler Riemannian theory to
a more complicated physical Lorentzian theory, which shall be used
in quantisation to relate the corresponding quantum theories. We will
call that type of phase space maps {\it internal}, because they leave
our real space-time base manifold unchanged, are restricted to the fibre
in our vector bundle formulation and substitute the signature therein.

\section{$2+1$ Wick transform} \label{Kap. $(2+1)$ Wick transform}

Let us now investigate the quantum Wick transform in $2+1$
Palatini theory. There are two classically equivalent formulations
of Lorentzian theory which naturally relate to Riemannian theory
and thus provide two different strategies to implement the
corresponding quantum Wick transform program. Our primary interest
is to study the particular appearance of the Wick transform
operator $\mathbf{\hat{W}}$ which will arise from the
\emph{internal Wick transform}, since this situation is analogous
to the one encountered for the Wick transform in $3+1$ gravity.
Here we investigate if appropriate internal phase space maps
between various $2+1$ theories can help to illuminate a deeper
geometrical nature of the quantum Wick transform
$\hat{\mathbf{W}}$ or even allow to extend our methodology to
simplify the appearance of canonical quantisation.

Its construction is completely analogous to the one in full gravity.
In Riemannian $2+1$ and $3+1$ Palatini theory we use the same type of
$\mathfrak{su}(2)$ valued connections and dyads or, respectively,
triads. Therein their constraints have identical structure. Thus,
the reader can easily translate the appropriate expressions from section
\ref{orig. W-Konzept von Abhay} to our $2+1$ dimensional setting.

We should stress here that the result of Wick transforming
$\mathfrak{su}(2)$ Riemannian theory is the $\mathfrak{su}(2)$
Lorentzian theory with complicated constraints. While this is a
good testing ground because of its similarity to 3+1 setting, we
don't have a good control over its solutions. We only understand
the solutions to the $\mathfrak{su}(1,1)$ formulation of the
Lorentzian theory. Recall that there, the reduced phase space is
the cotangent bundle over the moduli space of flat
$\mathfrak{su}(1,1)$ connections, which is finite dimensional.

Now having two classically equivalent Lorentzian theories, their
quantisation schemes should be comparable. The intention of our
further investigation is to understand complicated Wick transform
quantisation of $\mathfrak{su}(2)$ Lorentzian theory by relating
it to the simpler reduced phase space quantisation of its
equivalent $\mathfrak{su}(1,1)$ formulation. We do not control explicit
regularisation of our quantum Wick
transform $\hat{\mathbf{W}}$ in $\mathfrak{su}(2)$ Lorentzian
$2+1$ Palatini theory. However the existence of an alternative
$\mathfrak{su}(1,1)$ formulation for the physical Hilbert space of
Lorentzian theory should indirectly allow to find a regularisation
for $\hat{\mathbf{W}}$. By constructing an \emph{internal Wick rotation}
from the phase space of $\mathfrak{su}(2)$ Riemannian to $\mathfrak{su}(1,1)$
Lorentzian theory we will obtain a natural map from the moduli space of flat
$\mathfrak{su}(2)$ connections (i.e. reduced configuration space of Euclidean
theory) to the moduli space of flat $\mathfrak{su}(1,1)$ connections.
Its pull back to the algebras of functions on both quantum configuration
spaces provides a map from the Riemannian Hilbert space to the physical
Lorentzian Hilbert space. That construction is closely related to our Wick
transform and we will investigate if it represents an effective map on the
algebra of functions which in a more complicated way arises from
our Wick transform operator. The construction of such an
associated constraint preserving phase space rotation would
simplify the definition of our $2+1$ Wick transform operator
significantly and give hints for $3+1$ regularisation as well.

After our conceptual discussion which will be completed in section
\ref{Kap. $(2+1)$ Wick rotation} we will devote the last part of
this section to test explicit implementation of the quantum Wick
transform in $2+1$ gravity. Referring to the kinematical and
dynamical questions posed at the end of section \ref{orig.
W-Konzept von Abhay} we will concentrate on explicit
regularisation of the Wick transform operator $\mathbf{\hat{W}}$.
As a zeroth order regularisation test we will now investigate the
Wick transform operator not on the full kinematical Hilbert space
$\mathcal{H}_{kin}$, but instead under the following
simplifications. First, we simplify the appearance of our theories
by restricting our base manifold $\Sigma$ to be topologically
equivalent to the torus $T^2$. Second, we restrict
$\mathbf{\hat{W}}$ only to those functions in
$\mathcal{H}_{abel}\subset\mathcal{H}_{kin}$ which have domain in
a gauge fixed surface of the classical configuration space, which
is given by standard representatives \cite{Marolf Louko} from the
moduli space of flat $\mathfrak{su}(2)$ connections, i.e. we
restrict the quantum configuration space to Abelian connections
$A^I_a = (a_1d\varphi + a_2d\psi)\cdot t^I_3$ which can be
parameterised by the torus $(a_1, a_2)\in (0,4\pi ]\times (0,4\pi
] = T^2$ (see also section \ref{Space of classical solutions}).
Here, $\varphi$ and $\psi$ are the angular coordinates on $\Sigma$
and $t^{I}_{3}$ is a fixed basis vector of the internal vector
space $\mathfrak{su}(2)$. While restricting to
$\mathcal{H}_{abel}\subset\mathcal{H}_{kin}$ the momentum
operators $\hat{E}î_a$ arise from classical momentum variables
$E^I_a = (k_1d\varphi + k_2d\psi)\cdot t^I_3$ parameterised by
$(k_1, k_2)\in \mathbb{R}^2$. Thereon the Wick transform generator
$\hat{T}$ simplifies considerably and can alternatively be
understood to arise from the simplified classical generating
function $T = \frac{i\pi}{2}\int_{T^2} d^3x A^I_a
E^a_I$.\footnote{Under our gauge restriction for $(A^I_a, E^I_a)$,
the configuration variables of ADM - $K^I_a$, and connection
formulation - $A^I_a$, coincide due to vanishing Christoffel
connection $\Gamma = \Gamma(E^I_a)=0$.} The emerging picture for
the quantised kinematical phase space
$\mathcal{H}_{abel}\subset\mathcal{H}_{kin}$ and the Wick
transform thereon appears like the quantum theory which is
obtained from reduced phase space quantisation of $2+1$ gravity -
which also arises from the Abelian representative of the moduli
space of flat $\mathfrak{su}(2)$ connections (see section
\ref{Space of classical solutions}).

In particular we obtain $\mathcal{H}_{abel}=L^2(T^2)$ as the space
of functions on the quantum configuration space $T^2$ with
configuration operators $\hat{a}_i = a_i\cdot$ and momentum
operators $\hat{k}_i = i\hbar\frac{\partial}{\partial a_i}$.
Restricted to $\mathcal{H}_{abel}$ the Wick transform generator
appears as $\hat{T} = -\frac{\hbar\pi}{2}\frac{1}{2} \sum^2_{i=1}
a_i \frac{\partial}{\partial a_i} + \frac{\partial}{\partial a_i}
a_i $. The quantum Wick transform results in \bea
   \hat{\mathbf{W}} |_{\mathcal{H}_{abel}} & = & \exp \left( - \frac{1}{i\hbar} \hat{T}
   \right)= \exp \left( - i \frac{\pi}{4} \sum^2_{i=1}
   a_i \frac{\partial}{\partial a_i} + \frac{\partial}{\partial a_i}
   a_i \right)\\ & = & e^{ - i \frac{\pi}{4} (a_1 \frac{\partial}{\partial a_1}
   + \frac{\partial}{\partial a_1}
   a_1)} e^{ - i \frac{\pi}{4} (a_2 \frac{\partial}{\partial a_2} +
   \frac{\partial}{\partial a_2}
   a_2)} \label{Wtrans}
\eea which is the exponential of an anti self adjoint operator and
therefore well defined. Kinematically $\hat{\mathbf{W}}
|_{\mathcal{H}_{abel}}$ obviously maps constraint operators to
each other, since they are identically satisfied in both
Riemannian and Lorentzian description, i.e. we have
$\mathcal{H}_{abel}\subset\mathcal{H}_{phys}\subset\mathcal{H}_{kin}$.

Dynamically we further use that test example to get a glance on
the problem of finding an appropriate inner product in the
restricted part $\mathcal{H}_{abel}$ of the Lorentzian Hilbert
space. Due to vanishing Christoffel connection $\Gamma =0$ and
parameterisation of that gauge fixed part of the quantum
configuration space by the torus $T^2$ we can employ the same
canonical transformation $(a,k)\mapsto(\ln a,ak)$ which leads in
our Bianchi investigation to the Misner variables. Restricted to
$\mathcal{H}_{abel}\subset\mathcal{H}_{phys}$ the discussion of
$\hat{\mathbf{W}} |_{\mathcal{H}_{abel}}$ is similar. We again
realise, that in order to obtain normalised states from quantum
Wick transform, we have to restrict the Hilbert space of
Riemannian quantum theory by employing the same boundary
conditions which arose in our previous Wick transform
investigation on Bianchi cosmologies. Alternatively without the
canonical transformation $(a,k)\mapsto(\ln a,ak)$, for
completeness, in section \ref{explicit} we also provide explicit
construction of the action of the operator (\ref{Wtrans}) on
$\mathcal{H}_{abel}\subset\mathcal{H}_{phys}$ using the variables
$(a,k)$.

To summarise, by restricting $\hat{\mathbf{W}}
|_{\mathcal{H}_{abel}}$ to an invariant gauge fixed part of the
kinematical Hilbert space of the connection dynamical formulation
of $2+1$ gravity we simplify the generator $\hat{T}
|_{\mathcal{H}_{abel}}$, which now appears as a simple product of
canonically conjugate variables. For the Wick transform
quantisation program we obtain the same simplification and results
as for the corresponding utilisation of $\hat{\mathbf{W}}$ to
Bianchi models in the different framework of geometrodynamics.
However the task will be considerably harder to accomplish in full
$2+1$ gravity, because the structure of the generator $\hat{T}$ is
more complicated due to non vanishing Christoffel symbols and the
presence of an infinite dimensional quantum configuration space.

\section{$2+1$ Wick rotation} \label{Kap. $(2+1)$ Wick rotation}

As mentioned in the previous section, the physical states obtained from
the generalised Wick transform operator $\hat{\mathbf{W}}$ emerge in the
complicated $\mathfrak{su}(2)$ formulation. Alternatively, we will now use
the simpler $\mathfrak{su}(1,1)$ framework to test the Wick transform
quantisation indirectly by making use of the Wick rotation.

Let us now construct the \emph{internal Wick rotation}
$\mathfrak{W}$ between Riemannian and Lorentzian $2+1$ Palatini
theory. In both theories the phase space variables, are dyads and
pull backs of connection $1$-forms with values in either
$\mathfrak{su}(2)$ or $\mathfrak{su}(1,1)$. The map is obtained by
applying a $\mathfrak{sl}(2,\mathbb{C})$ rotation between
embeddings of both real theories into the complexified phase
space. Regarding utilisations of internal phase space isomorphisms
introduced before, this is an independent qualitatively new attempt
within $2+1$ gravity.

\subsection{Rotation in the complexified phase-space}\label{Rot. in C-phase-space}

The natural way to start is to notice that both Lie algebras are
identical after complexification
 \[
  \mathbb{C} \otimes \mathfrak{su}(2) = \mathbb{C} \otimes
  \mathfrak{su}(1,1) = \mathfrak{sl}(2,\mathbb{C})
 \]
The embedding in the complexified space is not unique. However,
the construction of the Wick rotation will unambiguously pick out
the required subspace identification.

Let us consider a basis $\mathrm{span} \{v_1,v_2,v_3\}_\mathbb{R}
\cong \mathfrak{su}(2)$ where the three positive norm base vectors obey
$\mathrm{span}\{v_1,v_2\}\perp v_3$ with respect to the
Cartan-Killing metric $\eta_{IJ}$. In order to utilise them as a
basis for $\mathrm{span} \{v_1,\ldots,iv_1,\ldots\}_\mathbb{R}
\cong \mathfrak{sl}(2,\mathbb{C})$ we choose linear extensions of
the structure constants $\varepsilon^{I}_{JK}$ and the
Cartan-Killing metric ${\eta_{IJ}}$ from $\mathfrak{su}(2)$ to
generalise them to unique objects on the complexified algebra,
denoted with the same symbols.

In order to approach $\mathfrak{su}(1,1)$ description, we
restrict these structures to the subspace: $\mathrm{span}
\{\ol{iv}_1,\ol{iv}_2,\ol{v}_3\}_\mathbb{R} \cong
\mathfrak{su}(1,1) \subset \mathfrak{sl}(2,\mathbb{C})$. This is
finally isomorphic to $\mathfrak{su}(1,1)$. Here, in order to
distinguish $\mathfrak{su}(2)$ valued one-forms from
$\mathfrak{su}(1,1)$ valued $1$-forms we use bars over the
$\mathfrak{su}(1,1)$ variables. We will also denote the
restriction of ${\eta_{IJ}}$ and $\varepsilon^{I}_{JK}$ to
$\mathfrak{su}(1,1)$ by bars.

Let us now extend the idea of that map to the phase-space of
Riemannian $2+1$ Palatini theory, parameterised by space-like
dyads and connection $1$-forms $(E^{I}_{a}, A^{I}_{a})$. We define
our internal Wick rotation to be completely determined by the
space-like dyads at every phase space point. Dyads define a unique
unit normal $n^{I}=n^{I}(E^{I}_{a})$ to the subspace spanned by
the dyads.\footnote{Independent of an arbitrary coordinate system
$(x_1, x_2)$ on the base manifold $\Sigma$ one can construct in a
gauge invariant way $n^{I}:=\frac{(E_{1} \times E_{2})^{I}}{\mid
(E_{1} \times E_{2})^I \mid} =
\frac{\varepsilon^{I}_{JK}E^{J}_{1}E^{K}_{2}}{\mid
\varepsilon^{I}_{JK}E^{J}_{1}E^{K}_{2} \mid}$.} Using that
decomposition of the internal vector space, the Cartan-Killing
metric decomposes into $\eta_{IJ}:=h_{IJ}+ n_{I}n_{J}$, where
$n_{I}$ is the dual to the space-like normal and $h_{IJ}$ is the
restriction of the internal metric $\eta_{IJ}$ to the subspace
spanned by the dyads.

To understand the complexification of our theory let us recall the
meaning of dyads as soldering forms between the tangent space
$T\Sigma$ and the internal vector space $\mathfrak{su}(2)$:
  \mbox{$E^{I}_{a} : \: T\Sigma \rightarrow \mathfrak{su}(2)$} and
with corresponding "inverse" $E^{a}_{I}$. In that sense
$iE^{I}_{a}$ represents a soldering form:
\[ iE^{I}_{a}:\;T\Sigma \longrightarrow i\:\mathfrak{su}(2) \subset
\mathfrak{su}(2) \oplus
i\:\mathfrak{su}(2) = \mathfrak{sl}(2,\mathbb{C}) \] Pulling back
$\eta^{\mathfrak{sl}(2,\mathbb{C})}_{IJ}$ respectively with
$E^{I}_{a}$ and $iE^{I}_{a}$, the signature of the induced metric
on $T\Sigma$ inverts. In particular one obtains $h_{ab}:=E^{\ast}
h_{IJ} = -(iE)^{\ast} \ol{h}_{IJ}=:-\ol{h}_{ab}$, emphasising the
nature of that Wick rotation. Completing the introduction of our
notation let us remark that $h^{I}_{J}:= E^{I}_{a}\circ E^{a}_{J}$
projects $\mathfrak{su}(2)$ to $\mathrm{span}
\{E^{I}_{1},E^{I}_{2}\}_\mathbb{R}$ irrespective of an arbitrary
coordinate system $(x_1,x_2)$ on $\Sigma$. The same follows for
 $\ol{h}^{I}_{J}:= \ol{iE}^{I}_{a}\circ \ol{iE}^{a}_{J}$, $\mathfrak{su}(1,1)$ and
 $\mathrm{span}\{\ol{iE}^{I}_{1},\ol{iE}^{I}_{2}\}_\mathbb{R}$.

To achieve our Wick rotation $\mathfrak{W}$, let us consider the
action of the following $\mathfrak{sl}(2,\mathbb{C})$ rotation on
a basis in the internal vector space:
\[
\begin{array}{cccc}
   \mathfrak{W} : & \mathfrak{su}(2) & \longrightarrow & \mathfrak{sl}(2,\mathbb{C}) \\
   \mathfrak{W} : & E^{I}_{1}, \: E^{I}_{2}, \: n^{I} & \longmapsto & \ol{iE}^{I}_{1},
   \: \ol{iE}^{I}_{2}, \: -\ol{n}^{I}
\end{array}
\]
Now we are ready to define the Wick rotation for an arbitrary
point $(E^{I}_{a}, A^{I}_{a})$ in the kinematical phase space of
Riemannian $2+1$ Palatini theory by expanding it to the
canonically conjugate connection $1$-forms:
\bea
   \mathfrak{W} (E^{I}_{a}) & := & i \cdot E^{I}_{a}
   \label{W-Definition-E}\\
   \mathfrak{W} (A^{I}_{a}) & := & i \cdot h^{I}_{J}A^{J}_{a} -
   n^{I}(n_{J}A^{J}_{a}) \label{W-Definition-A}
\eea
Here it is important to emphasise that the physically relevant
phase space variables are uniquely represented by connections. In
our vector bundle formulation they appear as covariant derivatives
$D_{a}=\partial_{a}+[A_{a},\cdot\:]$, which can be described
by reference covariant derivative ${\partial_{a}}$ and connection
$1$-form $A^{I}_{a}$. Therefore it is only legitimate to use
connection $1$-forms $A^{I}_{a}$ as phase space variables if we
define them with respect to a fixed reference covariant derivative
${\partial_{a}}$. For the detailed discussion of our phase space
variables from the viewpoint of differential geometry and the
associated distinction between connection $1$-forms $A^I_a$ and
the physically relevant connections we refer to appendix.
Hence to obtain a {\it well defined Wick
rotation} on connections we have to Wick rotate the reference
covariant derivative as well. For that purpose we choose
${\partial_{a}}$ to be flat and compatible with $n^I$ ($\partial_a
n^I=0$), i.e. depending on $E^I_a$ we have to fix an appropriate
$\partial_a=\partial_a (E)$ at every given phase space
point.\footnote{This does not include any gauge choice because we
simply decompose one and the same $D_a$ into a suitable reference
covariant derivative $\partial_a$ and the remaining connection
$1$-form $A^I_a$. Our Wick rotation of $D_a$ is independent of
that specific decomposition $D_a =
\partial_a + A_a$. For illustration see also example
(\ref{Pfirsich}).} This will allow us to treat
$\ol{\partial}_a:=\partial_a^{\mathfrak{sl}(2,\mathbb{C})}|_{
\mathfrak{W}(\mathfrak{su}(2))}$ as a flat $\mathfrak{su}(1,1)$
covariant derivative (see also appendix). Together with the associated connection
$1$-form $\mathfrak{W}(A^i_a)$ mapped according to
(\ref{W-Definition-A}) we define our Wick rotated
$\mathfrak{su}(1,1)$ covariant derivative as \bea
   \mathfrak{W}(D_a) & := &  \ol{\partial}_{a} + [\mathfrak{W}(A)_{a},
   \cdot\:] \: =: \: \ol{D}_{a}
\eea

The image of our Wick rotation lies in the phase space of
Lorentzian $2+1$ Palatini theory when we choose $-\ol{\eta}_{IJ}$
as our Lorentzian metric, where $\ol{\eta}_{IJ}:=\eta^{\mathfrak{sl}
(2,\mathbb{C})}_{IJ}|_{\mathfrak{W}(\mathfrak{su}(2))}$ and analogously
$-\ol{\varepsilon}^{I}_{JK}:=-(\varepsilon^{I}_{JK})^{\mathfrak{sl}
(2,\mathbb{C})}|_{\mathfrak{W}(\mathfrak{su}(2))}$ as our structure
constants of $\mathfrak{su}(1,1)$.
Consequently we can study the resulting theory in the independent
framework of real Lorentzian general relativity.

\subsection{From real Euclidean to real Lorentzian
$2+1$ gravity} \label{From real Eucl...}

So far the Wick rotation is just a map bringing together phase
spaces of two constrained Hamiltonian systems representing
Riemannian and Lorentzian $2+1$ gravity. Now we want to study how
this phase space map relates the "physical" content of both
theories.

Since we are dealing with first class constrained systems we draw
our attention to the symplectic structure
$\Omega(\delta_1,\delta_2) = \int[(\delta_1E^I_{a})
(\delta_2A^{aJ})-\leftrightarrow]h_{IJ}$ and the two constraint
functions: curvature constraint $\mathcal{F}^{I}_{ab} = F^I_{ab} = 0$ and
Gauss constraint $\mathcal{G}_{i} = D_{a}E^{a}_{i} = 0$. As
discussed in their initial presentation the only difference
between Lorentzian and Riemannian theory lies in the choice of
their internal Lie algebra. Meanwhile the structure of their
constraints is identical. Now we will analyse step by step how the
Wick rotation links the respective phase space structures of both
independent theories.

Let us first compare the curvature constraints $\mathcal{F}(E,A)$
and $\ol{\mathcal{F}}(\mathfrak{W}(E,A))$ which act on the phase
space of Riemannian and, respectively, Lorentzian theory if
related by our Wick rotation $\mathfrak{W}$. To compute the
curvature $\ol{\mathcal{F}}$ of the connection component at some
phase space point $\mathfrak{W}(E,A)$ we have to remember that
$\mathfrak{W}(A^I_a)$ is the connection $1$-form with respect to
the covariant derivative
$\ol{\partial}_a:=\partial_a^{\mathfrak{sl}(2,\mathbb{C})}|_{
\mathfrak{W}(\mathfrak{su}(2))}$ which was determined by the dyads
$E^I_a$. From appendix we know that $\ol{\partial}_a$ is again
flat and compatible with unit normal $n^I$ and internal metric
$\eta_{IJ}$. We obtain for the curvature in the emerging
Lorentzian theory:

\bea
   \ol{F}^{I}_{ab}(\mathfrak{W}(E,A)) & = & 2\ol{\partial}_{[a}
   (ih^{I}_{J}A^{J}_{b]})-2\ol{\partial}_{[a}(n^{I}n_{J}A^{J}_{b]}) \\
   & & - \ol{\varepsilon}^{I}_{JK}(ih^{J}_{L}A^{L}_{a} -
   n^{J}n_{L}A^{L}_{a})(ih^{K}_{L}A^{L}_{b} -
   n^{K}n_{L}A^{L}_{b}) \nn \\
   & = & i\cdot 2\partial_{[a}(h^{I}_{J}A^{J}_{b]}) -
   2\partial_{[a}(n^{I}n_{J}A^{J}_{b]}) \label{Paul1} \\
   & & -
   \ol{h}^{I}_{N}[ \ol{\varepsilon}^{N}_{JK}(ih^{J}_{L}A^{L}_{a} -
   n^{J}n_{L}A^{L}_{a})(ih^{K}_{L}A^{L}_{b} -
   n^{K}n_{L}A^{L}_{b})] \nn \\
   & & + n^{I}n_{N}[ \ol{\varepsilon}^{N}_{JK}(ih^{J}_{L}A^{L}_{a} -
   n^{J}n_{L}A^{L}_{a})(ih^{K}_{L}A^{L}_{b} -
   n^{K}n_{L}A^{L}_{b})] \nn \\
   & = & i\cdot 2\partial_{[a}(h^{I}_{J}A^{J}_{b]}) -
   2\partial_{[a}(n^{I}n_{J}A^{J}_{b]}) \label{Paul2} \\
   & & +
   \ol{h}^{I}_{N}[ \ol{\varepsilon}^{N}_{JK} (ih^{J}_{L}A^{L}_{a})
   (n^{K}n_{L}A^{L}_{b})+\ol{\varepsilon}^{N}_{JK}(n^{J}n_{L}A^{L}_{a})
   (ih^{K}_{L}A^{L}_{b})] \nn \\
   & & + n^{I}n_{N}[ \ol{\varepsilon}^{N}_{JK}(ih^{J}_{L}A^{L}_{a})
   (ih^{K}_{L}A^{L}_{b})]\nn \\
   & = & i\cdot 2\partial_{[a}(h^{I}_{J}A^{J}_{b]}) -
   2\partial_{[a}(n^{I}n_{J}A^{J}_{b]}) \label{Paul3} \\
   & & +
   \ol{h}^{I}_{N}[ i\varepsilon^{N}_{JK} (h^{J}_{L}A^{L}_{a})
   (n^{K}n_{L}A^{L}_{b})+i\varepsilon^{N}_{JK}(n^{J}n_{L}A^{L}_{a})
   (h^{K}_{L}A^{L}_{b})] \nn \\
   & & - n^{I}n_{N}[ \varepsilon^{N}_{JK}(h^{J}_{L}A^{L}_{a})
   (h^{K}_{L}A^{L}_{b})]\nn \\
   & = & i\cdot 2\partial_{[a}(h^{I}_{J}A^{J}_{b]}) -
   2\partial_{[a}(n^{I}n_{J}A^{J}_{b]}) \label{Paul4} \\
   & & +
   ih^{I}_{N}[ \varepsilon^{N}_{JK} (h^{J}_{L}A^{L}_{a})
   (n^{K}n_{L}A^{L}_{b})+\varepsilon^{N}_{JK}(n^{J}n_{L}A^{L}_{a})
   (h^{K}_{L}A^{L}_{b})] \nn \\
   & & - n^{I}n_{N}[ \varepsilon^{N}_{JK}(h^{J}_{L}A^{L}_{a})
   (h^{K}_{L}A^{L}_{b})]\nn \\
   & = & i\cdot h^{I}_{J} (2\partial_{[a}A^{J}_{b]}) -
   n^{I}n_{J}(2\partial_{[a}A^{J}_{b]}) \label{Paul5} \\
   & & +
   ih^{I}_{N}[ \varepsilon^{N}_{JK} A^{J}_{a}A^{K}_{b}] \nn \\
   & & - n^{I}n_{N} [\varepsilon^{N}_{JK}A^{J}_{a} A^{K}_{b}]\nn
\eea

In line (\ref{Paul1}) we used the projective identity map
$\ol{\delta}^{I}_{M}=\ol{h}^{I}_{M}-n^{I}n_{M}$ defined by using
our Lorentzian metric and commutativity of $\ol{\partial}$ with
$i$-multiplication. In line (\ref{Paul2}) the cross product
property of the structure constants. In step (\ref{Paul3})
and (\ref{Paul4}) we use that $\ol{\varepsilon}^{I}_{JK}$ and
$\ol{h}^{I}_{N}$ are just complex linear extension of
corresponding $\mathfrak{su}(2)$ objects. In (\ref{Paul5}) we use
again the cross product property of $\varepsilon^I_{JK}$ together
with an analogous $\mathfrak{su}(2)$ identity
$\delta^{I}_{M}=h^{I}_{M}+n^{I}n_{M}$ applied to the remaining
$\mathfrak{su}(2)$ terms and the compatibility of our flat
covariant derivative. Therefore we obtain:

\bea
   \ol{F}^{I}_{ab}(\mathfrak{W}(E,A)) & = & i\cdot h^{I}_{J} [2\partial_{[a}A^{J}_{b]}
   + \varepsilon^{J}_{KL} A^{K}_{a}A^{L}_{b}]  \\
   & & - \; n^{I}n_{J}[2\partial_{[a}A^{J}_{b]}+
   \varepsilon^{J}_{KL}A^{K}_{a} A^{L}_{b}] \; = \; \mathfrak{W}(F^{I}_{ab}(E,A)) \nn
\eea

Following the analogous line of arguments we will now evaluate the
Wick rotation with respect to Gauss constraint functions
$\mathcal{G}$ and $\ol{\mathcal{G}}$ in their particular domains.
Since all covariant derivatives are compatible with internal
metric $\eta_{IJ}$ we can alternatively compute
$\mathcal{G}^I:=D_{a}(E^{a I})$ :
\bea
   \ol{\mathcal{G}}(\mathfrak{W}(E,A)) & = & \ol{D}_{a}(\ol{i E^{a
   I}})\nn \\
   & = & \ol{\partial}_{a}iE^{a I} - \ol{\varepsilon}^{I}_{JK}(ih^J_M
   A^M_a - n^J n_M A^M_a)(iE^{aK})\nn \\
   & = & i\partial_{a}E^{a I} + \ol{h}^I_L[\ol{\varepsilon}^L_{JK}(n^J n_M
   A^M_a)(iE^{aK})]\nn \\
   & & \;\;\;\;\;\;\;\;\;\;\;\;- \; n^I n_L [-\ol{\varepsilon}^L_{JK}(ih^J_M
   A^M_a)(iE^{aK})]\nn \\
   & = & i\partial_{a}E^{a I} + i h^I_L[\varepsilon^L_{JK}
   A^J_a E^{aK}]\nn - n^I n_L [\varepsilon^L_{JK}
   A^J_a E^{aK}]\nn \\
   & = & i h^I_L[\partial_{a}E^{a L} + \varepsilon^L_{JK} A^J_a
   E^{aK}] - n^I n_L [\partial_{a}E^{a L} + \varepsilon^L_{JK} A^J_a
   E^{aK}]\nn\\
   & = & \mathfrak{W} (D_a E^{aI})\; = \;
   \mathfrak{W}(\mathcal{G}(E,A))\nn
\eea

Thus we observe phase space point by phase space point, that the
internal Wick rotation \textit{preserves all constraints}.

Furthermore by comparing the intrinsic symplectic structures of
both theories we realise that $\mathfrak{W}$ also preserves it.
The image of Riemannian phase space under $\mathfrak{W}$ will be
restricted non-holonomically to the part of Lorentzian phase space
with space-like dyad components only. Thereon the internal Wick
rotation provides a \textit{constraint preserving symplectic
isomorphism} from Riemannian to Lorentzian $2+1$ Palatini theory.
It maps constraint surface to constraint surface - especially flat
$\mathfrak{su}(2)$ connections to flat $\mathfrak{su}(1,1)$
connections.\footnote{We complete our local investigation of the
Wick rotation by emphasizing that so far these results hold only
restricted to open neighborhoods of \emph{non-degenerate} phase
space points, i.e. with non-degenerate dyads $E^I_a$. Otherwise
the Wick rotation is not defined.}

\subsection{Space of classical solutions} \label{Space of classical solutions}

Our aim is to discuss the effect of our map between "physical"
solution spaces of our Riemannian and Lorentzian constraint
systems. Globally this directs our attention to the effect of our
Wick rotation on the classically reduced phase space, appearing in
terms of gauge orbits in the constraint surface. Let us first
introduce the classical solution space of $2+1$ gravity.

Beginning in the full phase space of $2+1$ gravity our first class
constrained system is determined by Gauss $\mathcal{G}$ and
curvature $\mathcal{F}$ constraints. The reduced phase space of
classical solutions arises from symplectic reduction by first
restricting to the constraint surface and thereon identifying
gauge orbits generated by Hamiltonian vector fields corresponding to the constraints.
Points in the constraint surface will be given by pairs of flat
connections and respective dyad solutions to the Gauss constraint
$(A_{flat}, E_{Gauss})$. Within each orbit, the Gauss
constraint $\mathcal{G}$ generates internal $SU(2)$ or $SU(1,1)$
gauge transformations in the vector bundle. The curvature
constraint $\mathcal{F}$ generates gauge transformations on the
dyads $E^I_a$ only.

Thus one natural way to parameterise the reduced phase space is
obtained by choosing gauge fixed standard representatives
$(A_{flat}, E_{Gauss})$ in every gauge equivalence
class.\footnote{We parameterise the connection component of phase
space points by connection $1$-forms $A^i_a$ and reference
covariant derivative $\partial_a$. In contrast to our
$\mathfrak{W}$ definition, in the solutions space description we
further choose the gauge in our vector bundle formulation such,
that $\partial_a$ is represented by the partial derivative.}
Following Witten's connection formulation \cite{Witten} of $2+1$
gravity the pair $(A^I_a, E^I_a)$ of $\mathfrak{su}(2)$ connection
and dyad defines an $\mathfrak{isu}(2)$ connection in the trivial
principal bundle $\Sigma \; \times\; ISU(2)$. The constraints
$\mathcal{G}$ and $\mathcal{F}$ make the curvature of this
$\mathfrak{isu}(2)$ connection vanish. Further these constraints
generate local $ISU(2)$ gauge transformations. Thus the space of
classical solutions to our restated Riemannian $2+1$ gravity is
the moduli space of flat $\mathfrak{isu}(2)$ connections on
$\Sigma$, modulo $ISU(2)$ gauge transformations.

A flat connection on $\Sigma$ is completely determined by its
holonomies around closed non-contractible loops. The space of
classical solutions is therefore the space of group homomorphisms
from the fundamental group of $\Sigma$ to the holonomy group
$ISU(2)$, modulo conjugation by $ISU(2)$.
We choose the base manifold $\Sigma = T^2$. Here the space of
classical solutions is non trivial but qualitatively simpler than
for surfaces of genus two or higher, since the fundamental group
of the torus is the Abelian group $\mathbb{Z}\times\mathbb{Z}$.
Thus points in the space of classical solutions are just
equivalence classes of pairs of commuting elements of $ISU(2)$
under $ISU(2)$ conjugation, which represent holonomies along the
two fundamental loops on the torus.

Following the detailed implementation of Witten's connection
formulation on the torus $\Sigma = T^2$ as described by Louko and Marolf
in \cite{Marolf Louko} we obtain a unique gauge fixing on every gauge orbit in
the constraint surface. The standard representatives are given by
a pair $(A^I_a, E^I_a)$ of $\mathfrak{su}(2)$ connections
and dyads
\bea
\label{Marolf-Repraesentant}
 A^I_a & = & (a_1 d\varphi + a_2 d\psi)_{a} \cdot t^I_3 \\
 E^I_a & = & (k_1 d\varphi + k_2 d\psi)_{a} \cdot t^I_3 \nn
\eea
Here $(\varphi,\psi)$ is the pair of angular coordinates on the torus
and $a_{\alpha}$ and $k_{\alpha}$ are four parameters, with
$a_1,a_2\in(0,4\pi]$ and not both equal to zero. Also $\{t^I_1, t^I_2, t^I_3\}$
is a constant basis in $\mathfrak{su}(2)$.

Contrary to the Riemannian case, the moduli space of flat
$\mathfrak{isu}(1,1)$ connections
divides into three disconnected sectors. Let $\{t^{I}_{0},t^{I}_{1},t^{I}_{2}\}$
be a constant basis in $\mathfrak{su}(1,1)$, such that $t^{I}_{0}$
has a negative norm. Then, we obtain the following representatives
for the space-like sector:
\bea
\label{Marolf-Repr.-Lor-s}
 A^I_a & = & (a_1 d\varphi + a_2 d\psi)_{a} \cdot t^I_2 \\
 E^I_a & = & (k_1 d\varphi + k_2 d\psi)_{a} \cdot t^I_2 \nn
\eea
where the four parameters can take arbitrary values, except
that $a_{\alpha}$ are not both equal to zero. We obtain unique parameterisation
through the identification
$(a_{\alpha},k_{\alpha})\sim(-a_{\alpha},-k_{\alpha})$.
The time-like sector representatives are given by
\bea \label{Marolf-Repr.-Lor-t}
 A^I_a & = & (a_1 d\varphi + a_2 d\psi)_{a} \cdot t^I_0 \\
 E^I_a & = & (k_1 d\varphi + k_2 d\psi)_{a} \cdot t^I_0 \nn
\eea
where $a_{\alpha}$ and $k_{\alpha}$ are four parameters, with
$a_1,a_2\in(0,4\pi]$ and not both equal to zero. We will not be interested
in the null sector in what follows and we refer the reader to \cite{Marolf Louko}
or \cite{Bruno thesis} for more details.

 Now that we are familiar with the structure of the space of
classical solutions in Riemannian and Lorentzian $2+1$ gravity, let
us reclaim the Wick rotation. $\mathfrak{W}$ is so far only
defined on individual non-degenerate points in Riemannian phase
space. For the purpose of inducing our Wick rotation to the
classically reduced phase space we will, according to the
$\mathfrak{W}$ definition, use the explicit
description in terms of gauge orbits in the constraint surface.
Witten's connection formulation uniquely provides us in
(\ref{Marolf-Repraesentant}) with fixed $(A^I_a,E^I_a)$ on every
gauge orbit in the constraint surface. That representative, however,
has degenerate dyads and can not be Wick rotated.

 We will now specify a non-degenerate gauge fixed standard
representative $(A^I_a,E^I_a)$ for every gauge equivalence class
in the Riemannian solution space. To find the convenient gauge
fixing in the constraint surface we first select flat standard
connections $A_{flat}$ and supplement them with respective
standard Gauss constraint solutions $E_{Gauss}$. Afterwards we
prove their one to one correspondence to gauge orbits.

 Every flat $\mathfrak{su}(2)$ connection on the torus can be gauge
equivalently represented by an Abelian connection
\begin{equation}
\label{Abelzsh.}
 A^I_a = (a_1 d\varphi + a_2 d\psi)_{a} \cdot t^I_3
\end{equation}
with $a_1,a_2\in[0,4\pi)$ and not both equal to zero.
With respect to the constant basis $\{t^I_1, t^I_2, t^I_3\}$ in
the associated vector bundle our unrestricted ansatz for
respective Gauss constraint solutions is
\begin{equation} \label{dyad-ansatz}
 -\epsilon_{a}^{\; b} E_{b}^{I} =: \overset{\star}{}E^I_a =
 (f_1 d\varphi + g_1 d\psi)_{a} t^I_1 + \cdots +
 (f_3 d\varphi + g_3 d\psi)_{a} t^I_3
\end{equation}
where $f_1,\ldots,g_3$ are arbitrary functions on the torus. The Gauss
constraint $\mathcal{G}^I (E,A) = \epsilon^{ab}
D_{[a} \overset{\star}{}E^I_{b]}$ induces differential
equations to our dyad ansatz. One can show that the following
choice solves the equations (for details see \cite{Bruno thesis}).
We choose the constant solutions $f_1=a_1$, $g_1=a_2$
determined by our Abelian connection (\ref{Abelzsh.}), $f_2=g_2=0$
and $g_3=k_1$, $f_3=k_2$ with arbitrary constants $k_1,k_2$. Thus
we find our non-degenerate gauge fixed standard representative
$(A^I_a,E^I_a)$ for every gauge equivalence class
\begin{equation} \label{nicht-entart. Repr.}
\begin{array}{ccccc}
 A^I_a & = &  &  & (a_1 d\varphi + a_2 d\psi)_{a} \cdot t^I_3 \\
 E^I_a & = & (a_2 d\varphi + a_1 d\psi)_{a} \cdot t^I_1 & + & (k_1 d\varphi +
 k_2 d\psi)_{a} \cdot t^I_3
\end{array}
\end{equation}

These simple Gauss constraint solutions are sufficient to
represent all gauge orbits, since there is a one to one
correspondence between these non-degenerate representatives
(\ref{nicht-entart. Repr.}) and the degenerate representatives
(\ref{Marolf-Repraesentant}) arising from Witten's connection
formalism. Infinitesimal gauge transformations generated by the
curvature constraint act on phase space variables as
\[
\begin{array}{ccl}
 E^I_a & \longrightarrow & E^I_a \; + \; D^{(A)}_a\Lambda^I \\
 A^I_a & \longrightarrow & A^I_a \nn
\end{array}
\] where $\Lambda^I$ is a gauge field. The curvature gauge
transformation generated by $\Lambda^I = t^I_2$ maps our
degenerate representatives (\ref{Marolf-Repraesentant}) one to one
to its non-degenerate opponent (\ref{nicht-entart. Repr.}).

\subsection{Wick rotation in the solution space}

In order to investigate the Wick rotation restricted to the space
of solutions we will divide our consideratrions into three steps:
(i) First restrict to our non-degenerate gauge fixed
representatives and Wick rotate them individually. (ii) Thereafter
we extend $\mathfrak{W}$ to non-degenerate neighbourhoods in every
gauge orbit. (iii) Finally we will discuss what happens, when we
approach degenerate orbit points and draw conclusions to the
induced Wick rotation to full gauge orbits which include
degenerate dyads.

(i)   According to our definition in section \ref{Rot. in
C-phase-space} and remarks at (\ref{kov.Abl.4}) we will now Wick
rotate our non-degenerate gauge fixed phase space point
$(A^I_a,E^I_a)$ explicitly. This shall also illustrate the general
procedure for arbitrary phase space points. To Wick rotate the
connection component we decompose it suitably $A^I_a = Z^I_a +
\tilde{A}^I_a$ such that the corresponding covariant derivative
$D^{(A)}_a =
\partial_a + [A_a,\cdot]$ splits as follows
\begin{equation} \label{Pfirsich}
\begin{array}{ccccc}
   D^{(Z+\tilde{A})}_{a} & = & \underbrace{\partial_{a} +
   [Z_a,\cdot]} & + & [\tilde{A}_a,\cdot] \\
   & = & \partial^{(Z)}_{a} & + & [\tilde{A}_a,\cdot]
\end{array}
\end{equation} into a flat covariant derivative $\partial^{(Z)}_{a}$
compatible with $n^I$ ($\partial^{(Z)}_a n^I = 0$) and an
associated connection one-form $\tilde{A}^I_a$. Thereupon the
normal $n^I$ determines how to Wick rotate our reference covariant
derivative $\partial^{(Z)}_{a}$, the connection $1$-form
$\tilde{A}^I_a$ and the dyad $E^I_a$, what finally describes
$\mathfrak{W}(A^I_a,E^I_a)$.

Our non-degenerate phase space point (\ref{nicht-entart. Repr.})
was defined with respect to the orthonormal basis $\{t^I_1,
t^I_2,t^I_3\}$ in the associated $\mathfrak{su}(2)$ bundle. Using
angular coordinates $(\varphi, \psi)$ on the torus we realise that
the dyad components in (\ref{nicht-entart. Repr.}) span a
non-degenerate basis and define the normal $n^I=n^I(E)$ as \bea
 \begin{array}{c}
    E^I_{\varphi} = a_2 t^I_1 + k_1 t^I_3 \\
    E^I_{\psi} = a_1 t^I_1 + k_2 t^I_3
 \end{array}
 & \;\;\; \Longrightarrow \;\;\; & n^I = t^I_2
\eea By declaring our vector bundle basis $\{t^I_1, t^I_2,t^I_3\}$
to be parallel, we find a flat covariant derivative
$\partial^{(Z)}_{a}$ which is compatible with $n^I$. This makes
its defining connection to be horizontal, that is $Z^I_a=0$. Here
the representation of connections in terms of connection $1$-forms
is trivial, since we have $A^I_a = \tilde{A}^I_a$. We realise that
for our non-degenerate phase space points (\ref{nicht-entart.
Repr.}) the normal is always perpendicular to the connection
component $n^I \perp A^I_a$. According to definition
(\ref{W-Definition-A}), (\ref{W-Definition-E}) we therefore simply
Wick rotate $(A^I_a,E^I_a)$ as \begin{equation}
\label{W(A,E)-standard}
 \begin{array}{ccccr}
   \mathfrak{W} (A^{I}_{a}) & = & i \cdot A^{I}_{a} & = & (a_1 d\varphi +
   a_2 d\psi )_{a} \cdot \ol{i t}^I_3 \\
   \mathfrak{W} (E^{I}_{a}) & = & i \cdot E^{I}_{a} & = & (a_2 d\varphi + a_1 d\psi )_{a}
   \cdot \ol{i t}^I_1 + (k_1 d\varphi + k_2 d\psi )_{a} \cdot \ol{i t}^I_3
 \end{array}
\end{equation} These real $\mathfrak{su}(1,1)$ connections and dyads
$(\mathfrak{W}(A^I_a),\mathfrak{W}(E^I_a))$ are still embedded
into the complexified phase space. In accordance with our
$\mathfrak{W}$ definition at the end of section \ref{Rot. in
C-phase-space} the Lie algebra isomorphism
$(\mathfrak{W}(\mathfrak{su}(2)),-\ol{\varepsilon}^I_{JK}) \cong
(\mathfrak{su}(1,1), \varepsilon^I_{JK})$ allows us to identify
$ \mathfrak{W} \{t^I_1, t^I_2,t^I_3\} =
\{\ol{it}^I_1, \ol{t}^I_2,\ol{it}^I_3\}=:\{t^I_1, t^I_0,t^I_2\}$ as a
basis in $\mathfrak{su}(1,1)$.
Finally we obtain our Wick rotated non-degenerate phase space
point (\ref{nicht-entart. Repr.}) expressed in Lorentzian
$\mathfrak{su}(1,1)$ phase space variables as\footnote{The basis
notation is introduced in section \ref{Space of classical
solutions}. The vector bundle basis $\{t^I_1,t^I_2,t^I_3\}$
refers to the Riemannian phase space whereas
$\{t^I_0,t^I_1,t^I_2\}$ refers to the Lorentzian phase space.
Ambiguity will be avoided from the context.}
\begin{equation} \label{W(A,E)-imLor.Bild }
 \begin{array}{ccr}
   \mathfrak{W} (A^{I}_{a}) & = & (a_1 d\varphi + a_2 d\psi)_{a} \cdot t^I_2 \\
   \mathfrak{W} (E^{I}_{a}) & = & (a_2 d\varphi + a_1 d\psi)_{a}
   \cdot t^I_1 + (k_1 d\varphi + k_2 d\psi)_{a} \cdot t^I_2
 \end{array}
\end{equation}

 Let us range that particular rotation into the global solution
space structure. So far we individually Wick rotated for every
Riemannian solution space element the non-degenerate gauge fixed
orbit representative (\ref{nicht-entart. Repr.}) and obtained
associated gauge fixed points (\ref{W(A,E)-imLor.Bild }) in the
Lorentzian constraint surface. These $\mathfrak{W}(A^I_a,E^I_a)$
are all contained in the space-like sector of the Lorentzian
solution space. When we apply to $\mathfrak{W}(A^I_a,E^I_a)$ in
(\ref{W(A,E)-imLor.Bild }) the $\mathfrak{su}(1,1)$ curvature
constraint gauge transformation generated by the gauge field
$\Lambda^I=-t^I_0$ it is mapped to our degenerate gauge fixed
standard representative (\ref{Marolf-Repr.-Lor-s}) from the
space-like sector of Lorentzian solution space. Thus the gauge
fixed surface (\ref{nicht-entart. Repr.}) crossing all Riemannian
constraint orbits is Wick rotated to a gauge fixed surface
(\ref{W(A,E)-imLor.Bild }) which crosses some constraint orbits in
the space-like sector of Lorentzian phase space.

\begin{figure}
  \begin{center}
  \includegraphics[height=4cm]{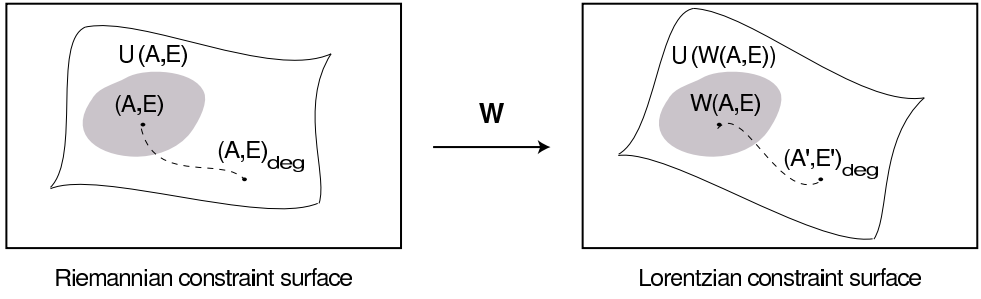}
  \caption{\small{Inside the Riemannian constraint surface we consider a constraint
orbit represented by $(A,E)_{deg}$ from (\ref{Marolf-Repraesentant}). Therein the
Wick rotation of non-degenerate representative $(A,E)$ from (\ref{nicht-entart.
Repr.}) shall be extended to its non-degenerate neighbourhood $\mathcal{U}(A,E)$.
In the Lorentzian constraint surface ${\mathcal{U}}(\mathfrak{W}(A,E))$
will belong to the constraint orbit represented by $(A',E')_{deg}$, which can not be
Wick rotated directly.}}
  \label{bruno1}
  \end{center}
\end{figure}

(ii) Let us now extend the point-wise Wick rotation
of gauge fixed orbit representatives to their local non-degenerate
neighbourhood in the associated constraint orbit. From local investigations
in section \ref{From real Eucl...} we know that our Wick rotation is
a constraint preserving, symplectic isomorphism. Therefore it locally maps
Riemannian phase-space flows generated by the constraints (i.e. gauge orbits)
onto the Lorentzian ones. Hence we conclude that $\mathfrak{W}$ is
locally compatible with non-degenerate parts of gauge orbits.

(iii) Consequently, connected non-degenerate part of a Riemannian
gauge orbit will be mapped onto connected part of a corresponding
Lorentzian gauge orbit.

\textit{If} the set of non-degenerate points in a constraint orbit
is \textit{simply connected}, i.e. two non-degenerate points can
always be connected by non-degenerate gauge trajectories, then all
these points will be Wick rotated into the same Lorentzian
constraint orbit. We \textit{would} obtain a unique map from
Riemannian to Lorentzian constraint orbits, by substitutional
mapping of respective non-degenerate orbit representatives in part
(i).

If the set of non-degenerate points in a Riemannian constraint
orbit is not simply connected, then there are parts separated by
degenerate boundary. The disconnected non-degenerate parts will be
individually mapped into their respective Lorentzian constraint
orbits. To obtain a global picture of the Wick rotated orbit, we
have to incorporate degenerate points at the boundary of our
non-degenerate parts for which $\mathfrak{W}$ is not defined and
see what happens to our Wick rotation when we cross them.
As discussed in \cite{Bruno thesis} there the infinitesimal
annihilation of dyad degeneracy in $(A,E)_{deg}$ to arbitrary
$n^I\in (E^I_{deg})^{\perp}$ and its implications on the global
continuity and orbit affiliation for our Wick rotation remain open
problems.

\section{Quantum Wick transform} \label{explicit}

The purpose of this section is to examine the action of the Wick
transform operator constructed in section \ref{Kap. $(2+1)$ Wick
transform}. We will explicitly construct its action on elements of
the Abelian Hilbert space $\mathcal{H}_{abel} \subset
\mathcal{H}_{phys}$, on the measure on that space as well as on a
complete set of operators which leave that space invariant. With
respect to the identification of
$\mathcal{H}_{abel}\simeq\mathcal{H}^{(red)}_{phys}$ with the
(equivalent!) reduced phase space quantisation our results can
formally be discussed as probing $\hat{\mathbf{W}}$ on physical
states and physical observables. Even though the Wick operator
provides the corresponding objects in the $SU(2)$ formulation of
Lorentzian theory, we will see that, with the use of Wick
rotation, we can naturally relate them to appropriate objects in
$SU(1,1)$ formulation.

 In order to evaluate the action of the Wick transform operator on
a state (which is a function of flat abelian connection $A_{0}$)
$\Psi^{E}(A_{0})\equiv\Psi^{E}(a_{1},a_{2})$, let us recall that
in the Euclidean theory $a_{1},a_{2} \in (0,4\pi]$ parameterise a
product of two circles $S^{1} \times S^{1}$. In the (spatial
sector of) Lorentzian theory, a state is also a function of two
real parameters. The difference is that now $a_{1},a_{2} \in
\mathbb{R}^{2}$ and the state has the property that
$\Psi(a_{1},a_{2})=\Psi(-a_{1},-a_{2})$. We can, however,
equivalently regard the Euclidean functions as periodic functions
on $\mathbb{R}^{2}$. This is what we will do from now on in order
to be able to map Euclidean states to the Lorentzian ones.
Therefore the Euclidean Hilbert space is now defined as the space
of square integrable functions on a 2-torus: $\mathcal{H}^{E} =
L^{2}(T^{2},{\rm d}\mu_{0})$, where the measure is given by ${\rm
d}\mu_{0} := {\rm d}a_{1} \wedge {\rm d}a_{2}$. It is well known
that such space is spanned by the basis which consists of
functions $\sin(n_{1}a_{1}+n_{2}a_{2})$ and
$\cos(m_{1}a_{1}+m_{2}a_{2})$, where $n_{I}$, $m_{I}$ ($I=1,2$)
are arbitrary integers. Each of the basis elements is well defined
not only for $(a_{1},a_{2}) \in \mathbb{R}^{2}$, but also for
arbitrary complex values of the two parameters. We will use this
property to evaluate the action of the Wick transform operator on
certain Euclidean states. We will use a dense subset of our
Hilbert space $\mathcal{H}^{E}$ for explicit evaluation. The
elements of the subset are finite linear combinations of $\sin$
and $\cos$ functions described above. For the rest of this section
$\Psi^{E}$ denotes a function which belongs to this dense subset.

 We can now make the following series expansion:
 \begin{equation}
  \Psi^{E}(a_{1},a_{2}) = \sum_{n=0}^{\infty} \frac{1}{n!} a_{2}^{n}
  \partial_{2}^{n} \Psi^{E}(a_{1},0)
 \end{equation}
Using this series expansion (and the analogous one in the other
variable), it is a straightforward calculation to show that:
 \begin{equation}
  \hat{\mathbf{W}} \Psi^{E}(a_{1},a_{2}) = {\rm e}^{i\frac{\pi}{2}}
  \Psi^{E}(ia_{1},ia_{2}) \label{WPsi}
 \end{equation}
It is worth making an interesting observation here that multiplication of the flat
connection by $i$ is precisely the action of the internal Wick rotation.
Indeed, by construction we have $n_{I} A_{0}^{I} =0$.\footnote{Recall that in order to
define $n^{I}$ (or Wick rotation) we need a non-degenerate frame $E^{I}_{a}$.
On the other hand, the frame used in our construction of $\hat{\mathbf{W}}$ and $\mathcal{H}_{abel}$
was degenerate (see section \ref{Kap. $(2+1)$ Wick transform}). In order to avoid
this problem we could, however, use non-degenerate representatives constructed in section
\ref{Space of classical solutions}. We would then obtain the same Hilbert space and the same
form of the operators $\hat{\mathbf{W}}$ and $\hat{T}$. Furthermore, notice that Wick rotation
restricted to those representatives is just given by multiplication by $i$. In fact,
it remains so for an arbitrary choice of the representative $E$.}
This also implies that $\mathfrak{W}(A_{0})$ belongs to the
space-like sector of $SU(1,1)$ moduli space. It should be stressed here,
that the above observation has been made for a very specific gauge fixed connection
and we don't know if it can be meaningfully extended to the whole gauge orbit.
Let us nevertheless use this formal relation of the Wick transform with the Wick rotation
to test its implications below. Thus, the state in the $SU(2)$ formulation of the Lorentzian
theory is now defined as $\Psi(a_{1},a_{2})$ such that
 \begin{equation}
  \hat{\mathbf{W}} \Psi^{E}(a_{1},a_{2}) = \Psi(a_{1},a_{2})
 \end{equation}
Note that (\ref{WPsi}) provides us with the following
formal prescription for a construction of $\Psi$.
Take a Euclidean solution $\Psi^{E}$, continue it analytically to
complexified phase-space and evaluate it on the Wick rotated
connection, i.e.
 \begin{equation}
  \Psi(A_{0}) = \Psi^{E}(\mathfrak{W}(A_{0})) \label{Wtr}
 \end{equation}
Note that \textit{quantum} Wick transform is described in terms of
\textit{classical} Wick rotation.

We can use it to map the restriction of physical observables to $\mathcal{H}_{abel}$
as proposed in \cite{Thiemann's W-paper,Abhay's W-paper}. As we know (see for example \cite{Abhay's Buch})
we can introduce the following, so called, loop variables which provide for us a
complete set of observables:
 \begin{gather}
  T^{0}[\alpha](A) := {\rm tr}h_{\alpha}(A) \\
  T^{1}[\alpha](A,E) := \oint_{\alpha} ds^{a} \epsilon_{ab}{\rm
  tr}E^{b}h_{\alpha}(A)
 \end{gather}
Those variables are labeled by a loop $\alpha \subset \Sigma$, $h_{\alpha}$
denotes the holonomy around the loop and the trace is taken in the fundamental
representation of the gauge group. When restricted to our reduced
(Euclidean)phase-space
$$T^{0}[n_{1},n_{2}](a_{1},a_{2})=2\cos(a_{1}n_{1}+a_{2}n_{2})$$
where $n_{1}$, $n_{2}$ are winding numbers of the loop $\alpha$. From
this, it is easy to show that
 \begin{equation}
  \hat{\mathbf{W}} \circ \hat{T}^{0}[n_{1},n_{2}](a_{1},a_{2}) \circ
  \hat{\mathbf{W}}^{-1} = 2 \cosh (a_{1}n_{1} + a_{2}n_{2})
 \end{equation}
which is indeed equal to the operator $T^{0}$ evaluated on
corresponding (through Wick rotation) $SU(1,1)$ connection.
Similarly $$T^{1}[n_{1},n_{2}](a_{1},a_{2},k_{1},k_{2})=
2(k_{1}n_{2}-k_{2}n_{1})\sin(a_{1}n_{1}+a_{2}n_{2})$$ Using
(\ref{Wtr}) one can then show that
 \begin{equation}
  \hat{\mathbf{W}} \circ \hat{T}^{1}[n_{1},n_{2}](a_{1},a_{2},k_{1},k_{2})
  \circ \hat{\mathbf{W}}^{-1} = 2i\hbar
  \sinh(a_{1}n_{1}+a_{2}n_{2})n_{I}\epsilon_{IJ}\frac{\partial}{\partial
  a_{J}}
 \end{equation}
which, again agrees with the operator appearing in the Lorentzian
$SU(1,1)$ theory.

 In order to discuss the scalar product, recall that on our reduced
quantum theory the appropriate product turns out to be
\cite{Abhay's Buch} $d\mu_{0}(A_{0}) = da_{1} \wedge da_{2}$. Thiemann
suggested in \cite{Thiemann's W-paper} a strategy for constructing a scalar
product on Wick transformed functions. The main condition for constructing
such a product was to provide a distribution $\nu$ on complexified
connections defined as:
 \begin{equation}
  \nu(A^{\mathbb{C}},\overline{A^{\mathbb{C}}}) = \left(
  \left(\overline{\hat{\mathbf{W}}^{\dagger}}\right)' \right)^{-1} \left(
  \overline{\left(\overline{\hat{\mathbf{W}}^{\dagger}}\right)'} \right)^{-1}
  \delta(A^{\mathbb{C}},\overline{A^{\mathbb{C}}})
 \end{equation}
where bar denotes complex conjugation, prime denotes analytic
continuation and the distribution $\delta$ is defined in
\cite{Thiemann's W-paper} (it reduces the integration over complex
connections to integration over the real section
$A^{\mathbb{C}}=\overline{A^{\mathbb{C}}}$). In our case of moduli
space of flat connections we can evaluate $\nu$ explicitly. Using
(\ref{WPsi}) we obtain:
 \begin{equation}
  \nu(A^{\mathbb{C}},\overline{A^{\mathbb{C}}}) \equiv
  \nu(a_{I}^{\mathbb{C}},\overline{a_{J}^{\mathbb{C}}}) =
  \delta(-ia_{I}^{\mathbb{C}},\overline{-ia_{J}^{\mathbb{C}}})
 \end{equation}
The result is therefore that, using our formal relation with Wick rotation,
the measure provided by Thiemann's construction restricts the integration over
complex connections to the real section which is exactly the $SU(1,1)$ section
defined by our Wick rotation. Furthermore, it reproduces the measure
$da_{1} \wedge da_{2}$ in which the Lorentzian observables
$\hat{T}^{0}$ and $\hat{T}^{1}$ are self-adjoint.

 Finally, we would like to ask if the Wick transform provides for us
sufficiently many Lorentzian states to be at all interesting. If we look at the
basis in $\mathcal{H}^{E}$ given by sin and cos functions as above,
the answer is not obvious. The reason for this is that the Wick
transform acting on any such basis element produces a sinh or cosh
function. The first of those does not belong to the Lorentzian physical Hilbert
space $\mathcal{H}^{L}$ and the second one is not normalisable on the real
line. Let us consider, however, a different subset of $\mathcal{H}^{E}$
 \begin{equation}
  \Psi^{E}_{n_{1},n_{2}}(a_{1},a_{2}) =
  f_{n_{1}}^{E}(a_{1})f_{n_{2}}^{E}(a_{2}) +
  f_{n_{1}}^{E}(-a_{1})f_{n_{2}}^{E}(-a_{2})
 \end{equation}
where $n_{1},n_{2}=0,1,2,3,\ldots$ and
 \begin{equation}
  f_{n}^{E}(a) := e^{(2n+1)ia} \exp(-\frac{1}{2}e^{2ia}) \label{Langley}
 \end{equation}
It is easy to see that under the Wick transform each function $f_{n}^{E}(a)$
is mapped to
 \begin{equation}
  f_{n}(a):=\hat{\mathbf{W}} f_{n}^{E}(a) = e^{-(2n+1)a}
  \exp(-\frac{1}{2}e^{-2a})
 \end{equation}
Since the result dies out (at least) exponentially when
$a\rightarrow\pm\infty$,
we obtain an element of the Lorentzian Hilbert space. In fact, we will
show that in this way we obtain a dense subset of
$\mathcal{H}^{L}$.\footnote{We would like to thank James Langley for
pointing out an example similar to (\ref{Langley}) and Jorma Louko
for the proof that this is a subset which is dense in $\mathcal{H}^{L}$.}
In order to do that, let us consider a family of functions
$g_{n}(x):=x^{n}e^{-\frac{1}{2}x}$, where
$x\in\mathbb{R}_{+}$ ($x$ is a positive real number). It is well known
that this is a complete set of functions in $L^{2}(\mathbb{R}_{+},dx)$.
By making the substitution $x=e^{-2a}$ we obtain functions on the real line
$g_{n}(a)=e^{-2na}\exp(-\frac{1}{2}e^{-2a})$. These form a complete set in
$L^{2}(\mathbb{R},e^{-2a}da)$. This, in turn, implies that
$f_{n}(a)=e^{-a}g_{n}(a)$ is a complete set in $L^{2}(\mathbb{R},da)$. Finally,
it is easy to see that the following set is complete in the Lorentzian
Hilbert space:
 \begin{equation}
  \Psi_{n_{1},n_{2}}(a_{1},a_{2}) =
  f_{n_{1}}(a_{1})f_{n_{2}}(a_{2}) +
  f_{n_{1}}(-a_{1})f_{n_{2}}(-a_{2})
 \end{equation}
Thus, we have shown that the quantum Wick transform is able to reproduce
all states of the Lorentzian theory (at least in this limited
context of $\mathcal{H}_{abel}$ in 2+1 gravity). Notice that this is not a
trivial statement, since the Wick transform is not an isomorphism!
Therefore, a priori, it was not known if it provides sufficiently
many (or any) Lorentzian states.

\section{Conclusions}

The generalised Wick transform quantisation method \cite{Thiemann's W-paper,
Abhay's W-paper} is an attractive strategy to simplify the appearance of
the last and the key step in the canonical quantisation program
for general relativity, namely solving the scalar constraint
$\hat{\mathcal{S}}_L$. In full quantum theory it has remained
formal. The main motivation behind this paper is to use simpler
truncated models to test if the program can be made rigorous and
to gain insight into the resulting physical states. Along with
these efforts there are three natural ways to implement the basic
Wick transform construction principle of employing \emph{internal}
isomorphisms in the complexified phase space to relate simpler
Riemannian theory with physical Lorentzian quantum theory:

(i) the \emph{usual Wick transform} relates different appearances
of the same Lorentzian theory, where we formally recover simpler
Riemannian constraints however subject to complicated reality
conditions.

(ii) in the \emph{generalised Wick transform} the previous concept
is lifted to the algebra of functions on the common real phase
space of both theories, thus providing a genuine map from real
Riemannian to real Lorentzian quantum theory.

(iii) the \emph{Wick rotation} is particularly adapted to the
simpler framework of $2+1$ gravity and employs a phase space
isomorphism between embeddings of both real theories into the
complexified phase space. In contrast to (i) it is a genuine phase
space map from Riemannian to Lorentzian classical general
relativity and naturally avoids reality conditions.

Our main emphasis is to test the particular quantisation program
which arises from the generalised Wick transform, in order to
explore associated subtleties which are expected to arise
similarly when applied to full quantum gravity. On \emph{Bianchi
models} the reduced Wick transform strategy can be implemented in
detail and is shown to reproduce the identical quantum theory as
obtained from Dirac quantisation. Furthermore we explicitly
observed for Bianchi type I, that to obtain physical states from
quantum Wick transform, one has to restrict the Hilbert space of
Riemannian quantum theory to states which are subject to
additional boundary conditions.

We investigate different connection dynamical formulations of
classical \emph{2+1 dimensional general relativity} and recover
among others the analog for the generalised Wick transform
therein. The corresponding Wick transform operator
$\mathbf{\hat{W}}$ causes similar regularisation problems as in
full gravity. Its restriction to a small subspace of Riemannian
quantum theory allows to study the operators explicitly and
glimpses to the same behaviour as on Bianchi models. However it
does not provide a full understanding for the quantum Wick
transform and the resulting physical states in $2+1$ gravity. This
is achieved by testing the Wick transform quantisation indirectly
in an appearance which classically arises from the internal Wick
rotation. Here we relate standard Riemannian theory to the simpler
$\mathfrak{su}(1,1)$ formulation of Lorentzian $2+1$ theory. When
induced to the reduced phase space this Wick transform provides a
natural map from solutions of Riemannian quantum theory to
physical states of Lorentzian quantum theory, however now
expressed in the simpler $\mathfrak{su}(1,1)$ formulation. Our
Wick rotation is a preliminary technique in $2+1$ gravity for our
main object of interest the generalised Wick transform
$\mathbf{\hat{W}}$. As presently formulated it can not be used to
regularise $\mathbf{\hat{W}}$ geometrically, however it allows to
characterise the resulting physical states. In the case of
space-time topology $M^{2,1}\cong T^2 \times \mathbb{R}$ we study
the resulting physical states explicitly and observe that the Wick
transform leads us to the most interesting, space-like sector of
$2+1$ gravity.

Fianlly, let us mention that the work done in section \ref{explicit}
can easily be extended to the $2+1$ gravity with a non-vanishing
cosmological constant. It is easy to check that Wick rotation and
Wick transform have the same properties when $\Lambda \neq 0$. Furthermore,
Ezawa has provided reduced phase-space representatives in \cite{Ezawa}.
As before, they can be chosen to be constant Abelian elements of the phase-space.
Because of that, the analysis of the section \ref{explicit} can be performed
without virtually any changes for arbitrary $\Lambda$.

\begin{acknowledgments} B.H. would like to thank the German National
Scholarship Foundation and the Pennsylvania State University for
financial support during one year stay at the Center for
Gravitational Physics and Geometry. We would like to thank Abhay
Ashtekar for the introduction into the research problem, important
suggestions and discussions. We also want to thank Martin Bojowald
for countless stimulating discussions. Finally, we would like to
thank Jorma Louko for the proof on the Lorentzian Hilbert space in
section VII and Thomas Thiemann for clarifying discussions.
\end{acknowledgments}

\appendix*

\section{Differential geometry of the phase space variables}\label{Diffgeo}

To understand the subtleties of our internal Wick rotation we
reflect the phase space variables from the perspective of
differential geometry \cite{Nakahara}. Their transparent
geometrical nature is usually hidden behind the coordinate
dependent terminology utilized in physics.

The mathematical space which carries all phase space variables of
connection dynamics is an adjoint vector bundle
$P\times_{Ad}\mathfrak{g}$, which is associated with a
$P(\Sigma,G)$ principal bundle. In our case $\mathfrak{g}$ is the
Lie algebra of $SU(2)$ or $SU(1,1)$. The dyads $E^{I}_{a}$
represent sections in the adjoint bundle $Ad\:P :=
P\times_{Ad}\mathfrak{g}$ and the connection $1$-forms $A^{I}_{a}$
are closely related to covariant derivatives $\partial^{(Z)}$
thereon.

On a principal fibre bundle $P$ a connection $Z\in\mathcal{C}(P)$
is a unique smooth right-invariant separation of the tangent space
$TP$ into vertical and horizontal subspace. Such a connection can
be described in terms of a $\mathfrak{g}$-valued $1$-form
$Z\in\Omega^1_{Ad}(P,\mathfrak{g})$ with the following
requirements:
\begin{align}
  (i) \; & Z(\Tilde{X})=X \;\;\;\;\;\; X\in\mathfrak{g} \nn \\
  (ii) \; & R^{\ast}_{g}Z=Ad_{g^{-1}}Z \nn
\end{align}
By adding a $1$-form
$\ol{A}\in\ol{\Omega}^1_{Ad}(P,\mathfrak{g})$, which is tensorial
(vanishes on vertical vector fields) and of type $Ad$ (under
right-multiplication with $G$) to a given connection $Z_1$ we
obtain another connection $Z_2:=Z_1+\ol{A}$. This makes the space
of all connections $\mathcal{C}(P)$ an affine space with the
vector space $\ol{\Omega}^1_{Ad}(P,\mathfrak{g})$. Further every
connection $Z$ uniquely defines an absolute differential
$D^Z:=d\circ\pi_{horiz}:\ol{\Omega}^k_{Ad}(P,\mathfrak{g})\rightarrow
\ol{\Omega}^{k+1}_{Ad}(P,\mathfrak{g})$, which satisfies:
\begin{equation}
   D^Z \ol{\omega} := d\ol{\omega} + Ad_{\ast}(Z)\wedge\ol{\omega} \label{kov.Abl.1}
\end{equation}

Our phase space variables act on the associated vector bundle
$Ad\:P := P\times_{Ad}\mathfrak{g}$. There exists a canonical
isomorphism: \bea
    \ol{\Omega}^k_{Ad}(P,\mathfrak{g}) & \Tilde{\longrightarrow} &
    \Omega^k(\Sigma,Ad\:P) \nn\\
    \ol{\omega} & \longmapsto & \omega:=[s,s^{\ast}\ol{\omega}]
    \label{ISO}
\eea where $[s,s^{\ast}\ol{\omega}]$ denotes the (gauge)
equivalence class in $Ad\:P$ which is obtained by choosing an
arbitrary section $s\in P$ and pulling back $\ol{\omega}$ from $P$
to $\Sigma$. For every connection $Z\in\mathcal{C}(P)$ this
isomorphism induces the absolute differential to a unique
covariant derivative $\partial^{Z}$ on the associated vector
bundle, which is defined on sections $[s,s^{\ast}\ol{\omega}]\in
\Omega^{0}(\Sigma,Ad\:P)$ as:
\begin{equation}
   \partial^{Z}\omega (X) := [s,s^{\ast}D^Z\ol{\omega}\:(X)] = [s,d\ol{\omega}(d s X) +
   Ad_{\ast}(Z (d s X)) \wedge \ol{\omega}\circ s] \label{kov.Abl.2}
\end{equation}
In particular for an arbitrary element $Z+\ol{A}$ in the affine
space of connections the corresponding unique element
$\partial^{Z+\ol{A}}$ in the affine space of covariant derivatives
can be parameterised in terms of $1$-forms $A\in
\Omega^1(\Sigma,Ad\:P)$ and a fixed covariant derivative
$\partial^{Z}$:
\begin{equation}
   \partial^{Z+\ol{A}}\omega(X) = \partial^{Z}\omega(X)+[s,Ad_{\ast}(s^{\ast}A(X))
   \wedge\ol{\omega}\circ s] \label{kov.Abl.3}
\end{equation}

Let us now recover our phase space variables in the more
coordinate dependent terminology in physics. All differential
forms in the associated vector bundle in (\ref{ISO}) are
equivalence classes and therefore defined depending on an
arbitrary section $s: \Sigma\rightarrow P$ in the principal
bundle, which is called \emph{gauge}. From now on we assume a
fixed gauge $s$ in $P$. The pull back $s^{\ast}\ol{\omega}$ of
$\ol{\omega}\in\ol{\Omega}^k_{Ad}(P,\mathfrak{g})$ is a
Lie-algebra valued k-form on the base manifold $\Sigma$. Following
these definitions all sections $\omega$ in (\ref{kov.Abl.2}) will
be described in terms of a $\mathfrak{g}$ valued function
$\omega^{I}:=\ol{\omega}\circ s$, pull backs of connection
$1$-forms $Z\in\mathcal{C}(P)$ in (\ref{kov.Abl.2}) will appear as
$\mathfrak{g}$ valued $1$-forms $Z^{I}_{a}:= s^{\ast}Z$ and pulled
back vector space elements $s^{\ast}A$ in (\ref{kov.Abl.3}) from
the affine space of connections will appear as $\mathfrak{g}$
valued $1$-forms $A^{I}_{a}:= s^{\ast}\ol{A}$. The adjoint
representation on the Lie algebra is nothing else than the
commutator $Ad_{\ast}A:=[A,\cdot]$ for $A\in\mathfrak{g}$. Let us
furthermore note, that all covariant derivatives in the adjoint
bundle $P\times_{Ad}\mathfrak{g}$ are naturally compatible with
$\eta_{IJ}$, which is induced from the $Ad$-invariant
Cartan-Killing metric on $\mathfrak{g}$.

The resulting formulation arises from suppressing the gauge in the
associated vector bundle formalism and from applying the above
abstract index notation. Following these lines the affine space of
all connections - i.e. the space of all covariant derivatives
(\ref{kov.Abl.3}) - can be parameterised by:\footnote{Only here
$\partial_a$ without index denotes the partial derivative which is
arising from the differential in (\ref{kov.Abl.1})}
\bea
   \partial^{(Z+\ol{A})}_{a} & = & \underbrace{\partial_{a}+Ad_{\ast}(Z^I_a)}
   +Ad_{\ast}(A^I_a)
   \label{kov.Abl.4}\\
   & = & \;\;\;\;\;\;\;\;\partial^{(Z)}_{a}\;\;\;\;\;\;\,+\varepsilon^I_{JK}(A^J_a) \nn
\eea
In particular its parameterization in terms of
\textit{connection $1$-forms} $A^{i}_{a}$ makes only sense with
respect to an once and for all fixed connection $Z^I_a$ - i.e.
fixed reference covariant derivative $\partial^{(Z)}_{a}$.

In the proposed internal Wick rotation (section \ref{Rot. in
C-phase-space}) the dyad component of the phase space variables
$(E,A)$ will determine how to map the connection. In our
parameterisation of the affine space of connections this will be
realised by simultaneously mapping \textit{connection $1$-forms}
$A^I_a$ and their reference covariant derivative $\partial_{a}$.
For that purpose we choose the reference $\partial_{a}$ in a way,
which allows us to identify $\ol{\partial}_a :=
\partial_{a}^{\mathfrak{sl}(2,\mathbb{C})}|_{
\mathfrak{W}(\mathfrak{su}(2))}$ as a suitable
$\mathfrak{su}(1,1)$ covariant derivative. This can be achieved
with every flat covariant derivative $\Tilde{\partial}_a$ which is
compatible with the normal $\Tilde{\partial}_an^I=0$ (and with
$\eta_{IJ}$ anyway).\footnote{ Consider the decomposition
$Ad\:P=Ad\:P_n \oplus Ad\:P_\perp$ with respect to $\eta_{IJ}$ and
$n^I=n^I(E)$ as it is defined in the Wick rotation. Because of
compatibility with $\eta_{IJ}$ we obtain for all sections
$u_{\perp}\in\Omega^0(\Sigma,Ad\:P_\perp)$ and
$u_{n}\in\Omega^0(\Sigma,Ad\:P_n)$ that $\Tilde{\partial}_a
u_{\perp} \in \Omega^1(\Sigma,Ad\:P_\perp)$ and
$\Tilde{\partial}_a u_{n} \in \Omega^1(\Sigma,Ad\:P_n)$.} Then, by
canonically extending $\Tilde{\partial}_a$ to the real
$6$-dimensional $\mathfrak{sl}(2,\mathbb{C})$ vector bundle its
restriction
$\ol{\Tilde{\partial}}_a:=\Tilde{\partial}_a^{\mathfrak{sl}
(2,\mathbb{C})}|_{ \mathfrak{W}(\mathfrak{su}(2))}$ is again flat.
We notice that $\ol{\Tilde{\partial}}_a \;i\cdot u_{\perp} =
i\cdot\Tilde{\partial}_a u_{\perp}$. For every non-degenerate dyad
$E^I_a$ such a $\Tilde{\partial}_a$ can be defined from extending
$n^I$ to a basis of parallel sections in the vector bundle.

\end{document}